\documentclass[12pt,english]{article}
\usepackage{natbib}
\usepackage{lmodern}

\usepackage[T2A,T1]{fontenc}
\usepackage[latin9]{inputenc}
\usepackage{geometry}
\usepackage{varioref}
\usepackage{xr}
\usepackage{float}
\usepackage{amsmath,mathrsfs,dsfont}
\usepackage{amsthm}
\usepackage{graphicx}
\usepackage{subcaption}
\usepackage{setspace}
\usepackage{systeme}
\usepackage{tikz}
\usetikzlibrary{positioning}

\makeatletter
\providecommand{\@fourthoffour}[4]{#4}
\def\fixstatement#1{%
  \AtEndEnvironment{#1}{%
    \xdef\pat@label{\expandafter\expandafter\expandafter
      \@fourthoffour\csname#1\endcsname\space\@currentlabel}}}

\newcounter{proofcount}

\long\def\proofatend#1\endproofatend{%
  \edef\next{\noexpand\begin{proof}[Proof of \pat@label]}%
  \toks\numexpr\prooftoks+\value{proofcount}\relax=\expandafter{\next#1\end{proof}}
  \stepcounter{proofcount}}

\def\printproofs{%
  \count@=\z@
  \loop
    \the\toks\numexpr\prooftoks+\count@\relax
    \ifnum\count@<\value{proofcount}%
    \advance\count@\@ne
  \repeat}
\makeatother

\usepackage{enumitem}

\usepackage{booktabs,caption}
\usepackage{verbatim} 

\usepackage{pdfpages}
\makeatletter

\usepackage{comment}

\usepackage[para,online,flushleft]{threeparttable}

\usepackage{graphicx}
\usepackage{wrapfig}
\usepackage{ amssymb }
\usepackage{wrapfig}
\usepackage{amsmath,amssymb}
\usepackage{bigints}
\usepackage{multirow}
\usepackage{stackrel}
\usepackage{setspace}
\usepackage{dcolumn}
\usepackage{caption}
\usepackage[nameinlink, capitalise, noabbrev]{cleveref}

\theoremstyle{plain}
\theoremstyle{plain}

\usepackage{pdflscape}
\usepackage{url}

\date{}
\usepackage{babel}

\makeatother

 {
      \theoremstyle{plain}
      
  }

\renewcommand{\ref}[1]{\hyperref[#1]{\ref{#1}}}

\newtheorem{proposition}{Proposition}
\newtheorem{lemma}{Lemma}
\newtheorem{corollary}{Corollary}

\renewcommand{\thefootnote}{\fnsymbol{footnote}}
\begin{document}

\title{\Large 
Competition, Politics, \& Social Media
}
\author{Benson Tsz Kin Leung\thanks{Faculty of Economics, University of Cambridge. Email: \href{mailto:btkl3@cam.ac.uk}{\texttt{btkl3@cam.ac.uk}}. }		\and	Pinar Yildirim\thanks{Corresponding Author: Assistant Professor, The Wharton School, University of Pennsylvania. 3730 Walnut St. Philadelphia PA 19104. Phone: (215) 746 2369.  Email: \href{mailto:pyild@wharton.upenn.edu}{\texttt{pyild@wharton.upenn.edu}}  }\footnote{Authors are in alphabethical order. We thank Christophe Van den Bulte, Yi Liu, and Vladimir Pavlov for their valuable feedback on the paper.} }
\maketitle

\begin{abstract}
An increasing number of politicians are relying on cheaper, easier to access technologies such as online social media platforms to communicate with their constituency. 
These platforms present a cheap and low--barrier channel of communication to politicians, potentially intensifying political competition by allowing many to enter political races. In this study, we demonstrate that lowering costs of communication, which allows many entrants to come into a competitive market, can strengthen an incumbent's position when the newcomers compete by providing more information to the voters. We show an asymmetric bad-news-good-news effect where early negative news hurts the challengers more than the positive news benefit them, such that in aggregate, an incumbent politician's chances of winning is higher with more entrants in the market. Our findings indicate that communication through social media and other platforms can intensify competition, however incumbency advantage may be strengthened rather than weakened as an outcome of higher number of entrants into a political market.  \\

\end{abstract}

	\renewcommand{\thefootnote}{\arabic{footnote}}
	\setcounter{footnote}{0}
\newpage


\setstretch{1.4}

\section{Introduction}

When does lowering barriers to entering a competition strengthen, as opposed to weaken, incumbency advantage? We study a special case of this broad question in the context of political races with barriers of marketing and communication. 
Historically, to reach out to large audiences, public personalities and brands mainly relied on expensive and limited formats of communication such as TV and newspaper advertising, or print and mailed pamphlets.  
Digital technologies altered much of that by exponentially reducing communication costs, reducing barriers for entrants into politics, and thus widening access. Online platforms became the preferred channel of communication for politicians to reach out to their constituency for campaigning and official communication. President Obama ran most of his 2012 campaign on Facebook \citep{borah2014facebook}, President Trump used Twitter as his official channel of communicating with the public while in office  \citep{kreis2017tweet}, and more than half of campaign advertising dollars in the 2020 Presidential race was spent on digital platforms \citep{Gibson2020}. 
On the one hand, social media communication enables more politicians, particularly newcomers who lack funds, to find a platform to make their voices heard. Via these cheaper technologies they can reach out to voters en masse and  inform them about their candidacy, values, policies, and campaign activities \citep{petrova2020social}. On the other hand, lowering barriers to entry can introduce a high number of challengers to an election who run not only against the incumbent, but also against {\em each other}. 
It is not ex ante clear whether lowering communication barriers to entering political races or higher intensity of information about newcomers helps or hurts incumbents. 

Incumbency advantage in politics is well-documented \citep[e.g.,][]{Ansolabehere06, Levitt97}. 75\% of senate races between 1980 and 2012 had a participating incumbent \citep{GarciaJimeno15}. These incumbents held a 1-2\% point advantage over their opponents in the 1940s, eventually widening up to a 8-10\% point advantage in early 2000s \citep{petrova2020social}. A host of factors contribute to the presence of a competitive advantage. Scholars suggest that incumbents hold a repeated advantage over their opponents \citep{Jacobson82}, likely because they are higher quality candidates \citep{ashworth2008electoral}, have access to resources of the offices they held, have more funding to run campaigns \citep{Cox96, fouirnaies2014financial}, and receive more media coverage compared to challengers \citep{Goldenberg80,Ansolabehere06, schaffner2006local}. The familiarity of voters with incumbent politicians and greater media coverage, combined with lack of funds to run costly advertising campaigns, erected tall barriers for newcomers who want to enter into political races, resulting in less competitive elections. Less competitive elections, in turn, result in worse economic, social, and democratic outcomes \citep{Myerson93,Persson03,Ferraz11,Galasso11}.

We build a model to study if lower informational barriers to enter into a race intensify competition, and in turn, reduce an incumbent's advantage. Focusing on a two-party political competition between differentiated candidates where informing and persuading voters is the end goal, we investigate if lower barriers of communication can alter the probability of winning for an incumbent politician. In our model, one party produces an incumbent as its candidate and the other produces one or more challengers on the opposing side. We incorporate incumbency advantage by allowing the incumbents to reach constituents more widely than the challengers and by allowing the constituents to have more positive prior beliefs about the incumbents, in line with \cite{ashworth2008electoral}. In this environment, we first investigate how additional challengers' entry to a race alters the incumbent's probability of winning the election and how this competitive advantage varies with the characteristics of the environment of competition, such as the effectiveness of the communication and advertising channels. 

We find that, while lowering barriers of marketing and communications {\em intensifies} political competition, it does {\em not} necessarily reduce the re-election probability of an incumbent. Lowering barriers to accessing communication channels can, in fact, {\em strengthen} the probability of winning for an incumbent when more challengers enter a race and when informational campaigns intensify. This is because, as more challengers enter a race, there is more media coverage, more communication via social media, and more campaign events by challenger candidates in the primaries\footnote{The observations from the 2020 Presidential Election in the US provide evidence for these statements. A record 29 challengers entered the race on the Democratic side \citep{Jacobson2019}. These candidates heavily utilized communication channels such as Facebook and Twitter, spent a record amount on advertising \citep{Fischer2020}, and held a high number of campaign events \citep{Schwartz2020}.} and these campaigns target or are followed by individuals who vote in the primaries.  
We show that more information arriving during the primary period has an asymmetric effect. 
Voters who receive a negative information (e.g., attack ads,  negative press coverage or social media buzz) about a challenger update their beliefs about the match of the candidate downward, which reduces the likelihood of voting for the candidate in the primary and in the general election against the incumbent. Similarly, a positive information received via the same means results in an upward update. The upside of a positive update, however, is small: while it increases the probability that a voter would support the candidate in the primary, it makes little difference in the chances of the candidate winning against the incumbent, since the voter is ex ante more likely to support the challenger against the incumbent, anyway. The potential harm from a negative news update, relative to the small upside of a positive news update, implies that more information during the primaries may hurt, rather than help a challenger win against an incumbent. 


We also analyze the impact of some recent policies in political communication space by online platforms such as Facebook, Google, and Twitter for narrowing options for political communication. Twitter eliminated political advertising on its platform entirely during 2020 \citep{Yaraghi2020}, Facebook reduced the ability to micro-target political ads and stopped political advertising a week before the 2020 U.S. Presidential Election \citep{Overly2020}, and Google similarly reduced micro-targeting for political advertisers \citep{Lee2019}. These bans were taken in an effort to reduce political tension and spread of misinformation in ads. They nevertheless reduce access to political information, and their consequences on electoral races have not been studied, to the best of our knowledge. We investigate the impact of these policies on the challengers' and the incumbents' ability to communicate with voters and argue that these bans may strengthen incumbency advantage and make elections less competitive.

Our study broadly contributes to the literature studying competition, entry barriers, strategic entry \citep[e.g.,][]{shen2010strategic, igami2014cannibalization, chen2018entry, joshi2009optimal} and incumbency advantage. \cite{demsetz1982barriers} notes that information costs are a fundamental barrier to entry, as they ``constitute hurdles to all who would (and have) enter(ed) the industry.'' Industrial organization literature, more specifically, recognizes advertising and promotions as informational barriers to entry \citep{demsetz1982barriers, schmalensee1983advertising}. Information can alter consumer tastes towards the advertised product and may erect additional challenges for the entrants \citep{cubbin1981advertising, dixit1978advertising, bagwell2007economic}. Advertising can also result in brand loyalty and consumer goodwill, thus new firms entering a market have to advertise more than the existing levels of advertising by incumbents to gain market share. Lowering informational barriers is particularly useful in markets with imperfect information, where there is sufficient uncertainty about the match value of products, and consumers resolve it through information received via advertising. This is also the case for political markets where new politicians with little-known policy positions frequently appear. 

Our paper contributes to this literature by explicitly focusing on the growing use of social and digital information channels, which lowers communication costs. While empirical marketing literature on social media has been growing \citep[e.g.][]{godes2009firm, schweidel2014listening}, there have been fewer theoretical examinations on the impact of social media channels \citep[e.g.,][]{joshi2015advertising, bart2017product}. 
To our knowledge, little focus is paid to easy access to media as an entry barrier. Recently  \cite{petrova2020social} studied if access to cheaper communication channels such as social media could earn politicians fundraising benefits. Authors find that, upon opening a social media account, an average politician's donations go up, but this increase is mainly observed for political newcomers rather than experienced politicians. Authors conclude that cheaper communication channels such as social media may mitigate incumbency advantage by allowing more politicians to enter into races and communicate with their constituency. Question is whether more communication facilitated by social media and cheaper digital communications, particularly that among the entrants which take place earlier in an election during the primaries, turn into a competitive advantage which can reverse the outcome of an election. Our study complements this paper by focusing on the voting outcome, using a theoretical model which incorporates an incumbent's informational advantage. This gives us a chance to address whether the incumbency advantage can be reversed. We demonstrate that, counter-intuitively, lowering communication barriers via the use of social media and digital advertising need {\em not} reduce incumbency advantage. An incumbent may preserve and increase his advantage with increasing number of challengers on the opposite side of the political spectrum.

Finally, we contribute to the literature documenting the long-standing incumbency advantage in the United States. 
Prior literature focused on the sources of incumbency advantage \citep{Levitt97}, listing structural advantages of being an incumbent as greater interest from media, fundraising, access to key individuals \citep{Ansolabehere00,Prat02,Stromberg04,Prior06, petrova2020social}. Incumbency advantage bars entry and reduces electoral competition, which in turn reduces accountability of politicians towards constituents \citep{Carson07}. More competitive elections result in better political and economic outcomes \citep{Myerson93,Persson03}.  Therefore understanding how lowering communication barriers can alter electoral competition is crucial.

In the rest of the paper, we first introduce the model in Section~\ref{sec: model} and follow with the analysis in Section~\ref{sec: analysis}. Section~\ref{sec:morethan2challengers} generalizes the model and offers extensions. We conclude in Section~\ref{sec: conclusion}.

\section{Model} \label{sec: model}

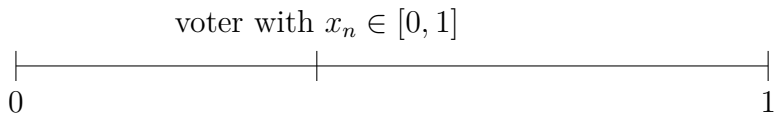
\begin{figure}[h]
    \centering
    \begin{tikzpicture}
    \draw (0,0)--(10,0);
    \draw (0,0.2)--(0,-0.2) node[below] {$0$};
    \draw (10,0.2)--(10,-0.2) node[below] {$1$};
    \draw (4,-0.2)--(4,0.2) node[above] {voter with $x_{n}\in[0,1]$};
    \end{tikzpicture}
    \caption{Hotelling line of political spectrum}
    \label{fig:hotelling}
\end{figure}

We develop a model of electoral competition, considering the race between an incumbent and one or more challengers on the opposite side of the political spectrum. Let the political ideology of voters be represented on a Hotelling line as illustrated in Figure~\ref{fig:hotelling}. Voters are represented by $n$ and their political ideology is denoted by $x_{n}$ and is uniformly distributed on a horizontal line, i.e., $x_{n}\sim U[0,1]$. Candidates are denoted by $i$ and without loss of generality, their ideology or political positions are assumed to be located at either end of this spectrum, on $0$ or $1$, labelled as ``left'' and ``right'' respectively. The political ideology of a candidate $i$ is assumed to be exogenous, and is denoted as $x_{i}\in\lbrace 0,1\rbrace$. Through the rest of the paper, for ease of exposition, we will refer to voters whose ideology is on the lower half of this spectrum ($x_n<1/2$) as left-wing and voters whose ideology is on the upper half of this spectrum ($x_n>1/2$) as right-wing voters.

Without loss of generality, we assume that the incumbent politician is a right-wing politician with $x_{i}=1$. Moreover, we assume that there can be at most two challenger candidates on the left hand side, taking an opposing left-wing position such that $x_{i}=0$.\footnote{We generalize the model to more than 2 challengers in Section~\ref{sec:morethan2challengers} in the paper.}   
We will refer to the incumbent candidate as candidate 3 ($i=3$) and the challengers as candidates 1 and 2 ($i=1,2$).

A voter's evaluation of electoral candidate $i$ is represented by $V(n,i)$ and depends on two factors: the ideological match between a candidate's and own political position and the individual evaluation of the candidate. Formally, 
\begin{equation}
V(n,i)=Q_{ni}-t(x_{n}-x_{i})^{2}+\epsilon\mathds{1}_{i=3}
\end{equation}
where $Q_{ni}$ is the personal evaluation of candidate $i$'s quality by voter $n$,  $-t(x_{n}-x_{i})^{2}$ is the distance between the voter's and candidate's political ideology, and $\epsilon\sim N(0,\sigma_{\epsilon}^{2})$ is a global taste shock of ideology that favors the incumbent politician iff $\epsilon>0$.\footnote{Alternatively, $\epsilon$ can also be interpreted as a piece of information received by all voters that could favor the left or right-wing candidate. This element is commonly used in probabilistic voting model \citep{lindbeck1987balanced}.} Here, $t$ measures the importance of the ideological match between the voter and a candidate. This modeling choice allows us to capture the similarities and differences between voters when they evaluate the same candidate. The ideological match component of the valuation ($-t(x_{n}-x_{i})^{2}$) allows a candidate to be valued similarly by voters of similar ideologies, while the idiosyncratic component $Q_{ni}$ allows a voter to prefer a candidate for reasons that are personal and independent of one's ideology. For instance, a voter may personally like the tax or education policy offered by a candidate and value the candidate more highly. 

Ex-ante, voters do not know the exact value of $Q_{n1},Q_{n2},Q_{n3}$, but know their prior distribution. They receive signals about the quality of the candidates during the primary and general election stages and update their beliefs, as will be described shortly.
We assume that voters' individual assessments about the quality of the incumbent is realized as a random draw from $Q_{n3}\sim N(q,\sigma_{Q}^{2})$ for some $q$ and that for a new candidate as a random draw from an unbiased distribution $Q_{ni}\sim N(0,\sigma_{Q}^{2})$ for $i=1,2$ and is independent across $n$.\footnote{The assumption of independence implies that $Q$ measures only the horizontal differentiation among the challengers but not the vertical differentiation. The model can be easily generalized to incorporate an element of vertical differentiation among the challengers, e.g., by adding a random variable $U$ that is common to all voter $n$. Our results hold as long as the upper bound of $U$ is not too large, or when the distribution of $U$ is not too dispersed.} Here
$q$ captures the difference in the expected quality between the incumbent and the entrants, which could be driven by the performance of the incumbent in his previous term. When $q>0$, the difference describes a form of incumbency advantage such that, on average, the incumbent candidate is assessed more positively. As discussed in the introduction, due to running political campaigns in the past or having been elected to an office, incumbents are known to hold a competitive advantage in elections over challengers \citep{Ansolabehere06b}. This incumbency advantage may be captured in the prior beliefs held.

\begin{figure}[ht]
    \centering
    \begin{tikzpicture}
	[auto,
	block/.style ={rectangle, draw=black, thick, text width=11em,align=center, rounded corners, minimum height=1.5em}]
    \node[block] (start) {\footnotesize Candidates $1$ and $2$ decide whether to enter the race at a cost $C$.};
    \node[block, below left = 3em and 3em of start] (0a) {\footnotesize Both do not enter.};
    \node[block, below = 3em of 0a] (0b) {\footnotesize Only candidate $3$\\\footnotesize participates and wins in the general election.};
    \node[block, below =3em of start] (1a) {\footnotesize Only candidate $i\in\lbrace1,2\rbrace$\\\footnotesize enters.};
    \node[block, below right = 3em and 3em of start] (2a) {\footnotesize Both candidates $i=1,2$\\\footnotesize enter.};
    \node[block, below = 3em of 2a] (2b) {\footnotesize Voters with $x_{n}<\frac{1}{2}$\\\footnotesize acquire information about candidates $1$ and $2$ and vote in the primary election.};
    \node[block, below = 3em of 2b] (2c) {\footnotesize Candidate $i\in\lbrace1,2\rbrace$ wins the primary election.};
    \node[block, below left= 3em and -5em of 2c] (2d) {\footnotesize Voters acquire information about candidates $i$ and $3$ and vote in the general election after observing $\epsilon.$ Candidate with the highest support wins. };
    \draw[thick,->] 
	(start)--(0a);
	 \draw[thick,->] 
	(start)--(1a);
	    \draw[thick,->]
	(start)--(2a);
	    \draw[thick,->] 
	(0a)--(0b);
	    \draw[thick,->] 
	(2a)--(2b);
	    \draw[thick,->] 
	(2b)--(2c);
	    \draw[thick,->] 
	(2c) |- (2d);
	    \draw[thick,->] 
	(1a) |- (2d);
    \end{tikzpicture}
    \caption{Timeline of the game}
    \label{fig:timeline}
\end{figure}
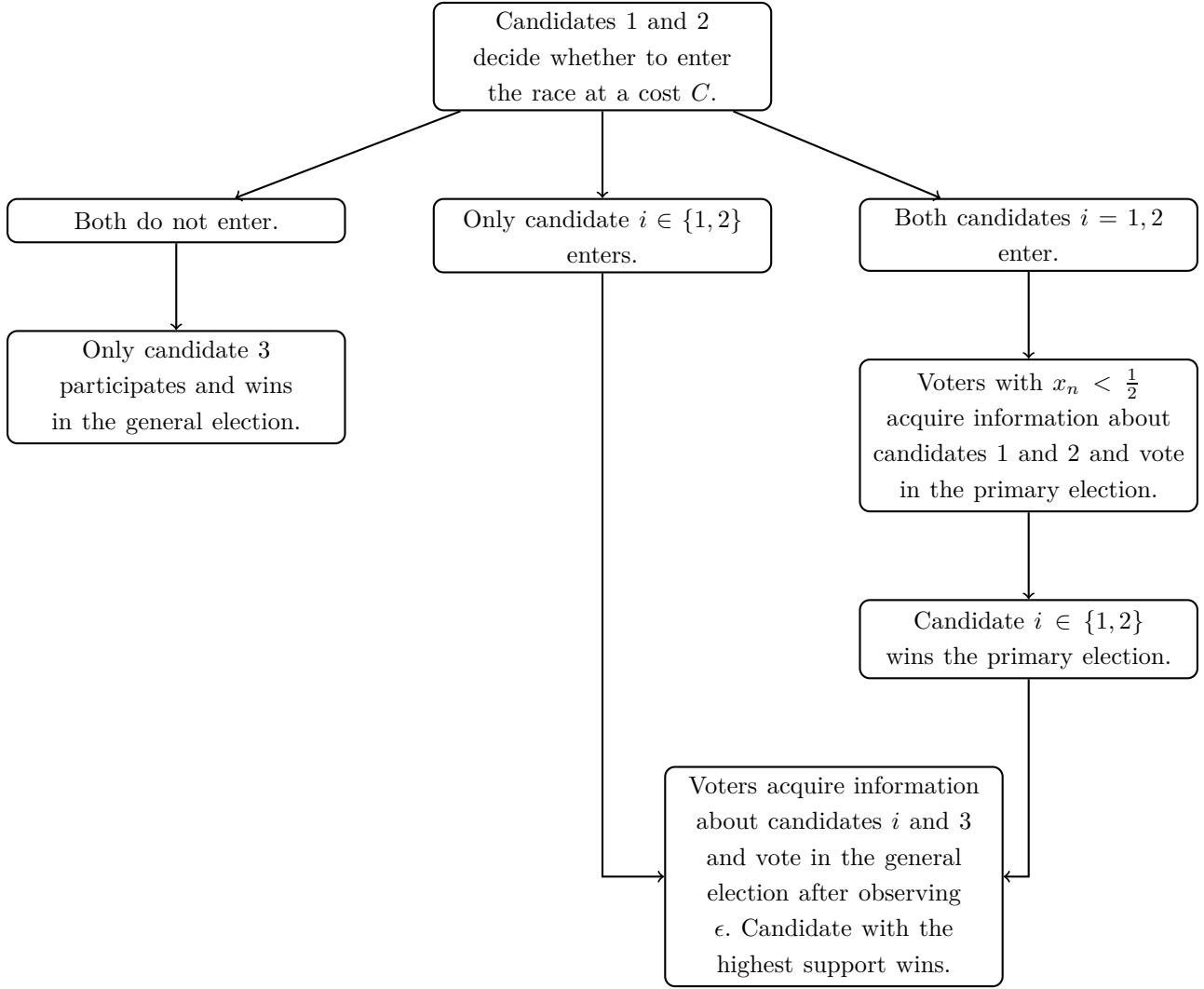

The game timeline is illustrated in Figure~\ref{fig:timeline} and lasts two periods, with a possible primary stage and a general election stage. In period~$1$, the challengers (candidates $1$ and $2$) decide whether to enter the election at some fixed cost $C$.\footnote{The election setting we consider resembles a senate election in the U.S., where the race has (historically) been between two party candidates, there is a primary period leading to the elections, and the winner is elected by plurality of votes.} This cost represents, in our framework, the barrier to entering a market. While there can be a number of such barriers, in line with the focus of our paper, we will treat this cost as the cost of communication. If there are no challengers or if only one challenger enters the race, a primary election is not necessary to determine the candidate for the general election from the left, and the game goes straight to the general election. If challengers $1$ and $2$ both decide to enter, however, they compete in a primary.\footnote{Notice that, technically, there is also a primary on the side of the incumbent. We abstract away from modeling a challenger on the side of the incumbent for simplicity, but the solutions which involve a challenger on the incumbent's side can be obtained from the authors. The key qualitative insights of the paper are not altered by this modification. The model with a single incumbent captures the competitive advantage of the incumbent in a parsimonious model.}

Note that, in a primary, only the voters with left-wing views, $x_{n}<\frac{1}{2}$, vote and they learn about the quality of the challengers by gathering noisy information about them. They then vote for the candidate with the higher expected value. More specifically, we assume that the communications during the primary are aimed at the voters who participate in the primaries, such as advertising or social media messages of the candidates.  
Each voter $x_{n}<\frac{1}{2}$ receives a private signal $s^{p}_{ni}$ about candidate $i=1,2$ from such communications, where $p$ stands for `primary.' Signals $s^{p}_{n1}$ and $s^{p}_{n2}$ are independently distributed according to $N(Q_{ni},\sigma_{s}^{2})$.

After receiving signals $s^{p}_{n1},s^{p}_{n2}$, voters $n$ with political ideology $x_{n}<\frac{1}{2}$ update their expected valuation of candidates 1 and 2 and vote for the one with the highest expected value in the primary election. More specifically, they update their expected value according to the Bayes' rule, as follows:
\begin{equation}
    \begin{split}
        E(V(n,1)\mid s^{p}_{n1})&=\frac{\sigma_{s}^{2}}{\sigma_{Q}^{2}+\sigma_{s}^{2}}E(V(n,1))+\frac{\sigma_{Q}^{2}}{\sigma_{Q}^{2}+\sigma_{s}^{2}}s^{p}_{n1}-tx_{n}^{2}=\frac{\sigma_{Q}^{2}}{\sigma_{Q}^{2}+\sigma_{s}^{2}}s^{p}_{n1}-tx_{n}^{2},\\
        E(V(n,2)\mid s^{p}_{n2})&=\frac{\sigma_{s}^{2}}{\sigma_{Q}^{2}+\sigma_{s}^{2}}E(V(n,2))+\frac{\sigma_{Q}^{2}}{\sigma_{Q}^{2}+\sigma_{s}^{2}}s^{p}_{n2}-tx_{n}^{2}=\frac{\sigma_{Q}^{2}}{\sigma_{Q}^{2}+\sigma_{s}^{2}}s^{p}_{n2}-tx_{n}^{2}.
    \end{split}
\end{equation}

Note that, during the primary stage, voters with $x_n<\frac{1}{2}$ receive a signal about each challenger candidate. Importantly, voters with $x_{n}>\frac{1}{2}$ do not vote in the left-wing primary, and therefore receive less information, which we normalize to no signals in the model for simplicity. This asymmetry reflects the fact that voters with aligned political ideology pay more attention to the primary compared to voters whose ideology is misaligned with the party whose primary is held. 

Next, suppose candidate $i\in\lbrace1,2\rbrace$ participates in the general election, either via winning the primary election or due to being the only candidate on the left. In the general election, voters receive extra information about the candidate $i\in\lbrace1,2\rbrace$ on the left and candidate $3$ on the right. In general election, all voters receive a signal $s_{n3}^{g}$ about the incumbent that follows $N(Q_{n3},\beta\sigma_{s}^{2})$, where $g$ stands for general election. 
On the other hand, voters with $x_{n}<\frac{1}{2}$ receive a signal $s_{ni}^{g}$ about the challenger candidate $i\in\lbrace 1,2\rbrace$ which follows $N(Q_{ni},\beta\sigma_{s}^{2})$, while voters with $x_{n}>\frac{1}{2}$ receive a signal $s_{ni}^{g}$ about the challenger candidate $i$ which follows $N(Q_{ni},\lambda\beta\sigma_{s}^{2})$.

Here, two parameters allow us to study the differentiation between the effectiveness of political marketing campaigns during the primary and general election stages, and that between the incumbent and the challengers. They are not, however, necessary for driving our general qualitative insights.

First, we introduce 
$\beta\in(0,\infty)$, which measures the informativeness of the signals in a general election compared to the primary election for all candidates. Specifically, this parameter allows the general election marketing campaigns to yield more or less precise signals in informing voters relative to the primary stage. It is possible, for instance, that politicians use different formats of communication or advertising, or media or voters pay more attention to the election during the primary stage. Or as we will show later, more competition could lead to more activities and information provision in the primary. All of these differences between the primary and general election environment that are common across the politicians would be captured by this parameter, and a larger (smaller) $\beta$ indicates a more (less) informative signal about the candidates in the primary period compared to that in the general election. 

Second, the parameter $\lambda\geq 1$ captures any residual disadvantages that challengers face due to being lesser-known relative to the incumbent among the right-wing supporters after the primary period. When $\lambda=1$, there are no disadvantages. When $\lambda>1$, the challengers' signals are less effective in informing the supporters of the incumbent ($x_n>1/2$). 
We introduce this parameter to capture the challenges faced by entrants during a general election when they are trying to communicate with individuals whose ideological positions are not aligned with theirs. 
This is motivated by the fact that voters with $x_n>1/2$ do not vote in the primary period, thus they are less likely to pay attention to a primary that they are not voting in relative to those who do. For instance, they are less likely to be readily available in campaign events, follow news about  challengers or follow them on social media. As a result, those who do not vote in the primary are less susceptible to information from challengers and they may have to rely on paid communication tools such as advertising to reach out to these voters. 
The parameter $\lambda$ allows us to capture the difference between the incumbent and challengers in the ability to reach out to incumbent's base (voters with $x_n>1/2$) using communication tools. Restrictions specific to paid political communication, for instance, narrows a challenger's opportunity to reach voters with $x_n>1/2$. As reported in the literature, for incumbents, informing voters with $x_n<1/2$ is less an issue because of having been elected to an office before and the ample media coverage they receive \citep{Ansolabehere06}. 
In the period leading to the 2020 U.S. Presidential Election, Twitter banned political advertising on its platform \citep{Yaraghi2020}, Facebook banned political ads in the week leading to the election, and Google limited political advertisers' micro-targeting ability \citep{Lee2019}. These changes, for instance, may narrow a challenger's ability to reach out to a broad set of voters (i.e., result in a higher $\lambda$).

To summarize the information provision in a general election, upon receiving $s_{ni}^{g}$, $i\in\lbrace 1,2\rbrace$ and $s_{n3}^{g}$, voters with $x_{n}>\frac{1}{2}$ update their expected value from voting for each candidate according to:
\begin{equation}
    \begin{split}
        E(V(n,i)\mid s^{g}_{ni})&=\frac{\sigma_{Q}^{2}}{\sigma_{Q}^{2}+\lambda\beta\sigma_{s}^{2}}s^{g}_{ni}-tx_{n}^{2};\\
        E(V(n,3)\mid s^{g}_{n3})&=\frac{\beta\sigma_{s}^{2}}{\sigma_{Q}^{2}+\beta\sigma_{s}^{2}}q+\frac{\sigma_{Q}^{2}}{\sigma_{Q}^{2}+\beta\sigma_{s}^{2}}s^{g}_{n2}-t(x_{n}-1)^{2}.
    \end{split}
\end{equation}
On the other hand, if there was no primary preceding the general election, upon receiving $s_{ni}^{g}$, $i\in\lbrace 1,2\rbrace$, and $s_{n3}^{g}$ in the general election, voters with $x_{n}<\frac{1}{2}$ update their expected valuation according to:
\begin{equation}
    \begin{split}
        E(V(n,i)\mid s^{g}_{ni})&=\frac{\sigma_{Q}^{2}}{\sigma_{Q}^{2}+\beta\sigma_{s}^{2}}s^{g}_{ni}-tx_{n}^{2};\\
        E(V(n,3)\mid s^{g}_{n3})&=\frac{\beta\sigma_{s}^{2}}{\sigma_{Q}^{2}+\beta\sigma_{s}^{2}}q+\frac{\sigma_{Q}^{2}}{\sigma_{Q}^{2}+\beta\sigma_{s}^{2}}s^{g}_{n2}-t(x_{n}-1)^{2}.
    \end{split}
\end{equation}
If there was a primary election preceding the general election, upon receiving $s_{ni}^{g}$, $i\in\lbrace 1,2\rbrace$, and $s_{n3}^{g}$ in the general election, voters with $x_{n}<\frac{1}{2}$ update their expected valuation according to:
\begin{equation}
    \begin{split}
        E(V(n,i)\mid s^{p}_{ni},s^{g}_{ni})&=\frac{\frac{1}{\beta\sigma_{s}^{2}}}{\frac{1}{\sigma_{Q}^{2}}+\frac{1}{\beta\sigma_{s}^{2}}+\frac{1}{\sigma_{s}^{2}}}s^{g}_{ni}+\frac{\frac{1}{\sigma_{s}^{2}}}{\frac{1}{\sigma_{Q}^{2}}+\frac{1}{\beta\sigma_{s}^{2}}+\frac{1}{\sigma_{s}^{2}}}s^{p}_{ni}-tx_{n}^{2};\\
        E(V(n,3)\mid s^{g}_{n3})&=\frac{\beta\sigma_{s}^{2}}{\sigma_{Q}^{2}+\beta\sigma_{s}^{2}}q+\frac{\sigma_{Q}^{2}}{\sigma_{Q}^{2}+\beta\sigma_{s}^{2}}s^{g}_{n2}-t(x_{n}-1)^{2}.
    \end{split}
\end{equation}

After both stages, and having observed the realization of $\epsilon$, voters vote for the candidate which generates higher expected value for them in the general election. The candidate who wins the general election is assumed to receive a utility or a prize of $1$. If no candidates on the left enter the election, candidate $3$ wins with probability $1$.

\section{Analysis} \label{sec: analysis}
We next solve the game described in the previous section using backward induction. To this end, we will consider each sub-game of a general election where there can be zero, one, or two challenger candidates entering the race and then we will compare these cases to assess if more challengers reduce incumbency advantage in a general election. We will also draw general insights about the informational disadvantages of the challengers and how they contribute to the presence of the incumbency advantage.  

\paragraph{No challengers enter the election.} We first consider the case when there are no challengers. The analysis of this case is straightforward since candidate $3$ wins the election  and gets payoff $1$, by default, and candidates $1$ and $2$ receive $0$ payoff.

\paragraph{Only one challenger enters the election.}
Next, we consider the case when there is only one challenger entering the race. Without loss of generality, we will assume that candidate $1$ enters the race as the challenger on the left. The expected valuation of voter $x_{n}$ of candidates 1 and 3, conditional on receiving $s^{g}_{n1}$ and $s^{g}_{n3}$, are now  
\begin{equation}
    \begin{split}
        E(V(n,1)\mid s^{g}_{n1})&=\frac{\sigma_{Q}^{2}}{\sigma_{Q}^{2}+\lambda'\beta\sigma_{s}^{2}}s^{g}_{n1}-tx_{n}^{2},\\
        E(V(n,3)\mid s^{g}_{n3})&=\frac{\beta\sigma_{s}^{2}}{\sigma_{Q}^{2}+\beta\sigma_{s}^{2}}q+\frac{\sigma_{Q}^{2}}{\sigma_{Q}^{2}+\beta\sigma_{s}^{2}}s^{g}_{n3}-t(x_{n}-1)^{2}+\epsilon
    \end{split}
    \label{eq:evaluationonecandidate}
\end{equation}
where $\lambda'=1$ for voters with $x_{n}<\frac{1}{2}$ and $\lambda'=\lambda$ for voters with $x_{n}>\frac{1}{2}$. In the following, we characterize the vote received by candidate $3$ given $\epsilon$. Fixing $\epsilon$, a voter with political ideology $x_{n}$ votes for candidate $3$ if and only if upon updating the beliefs about the candidate, he has a higher expected valuation, which holds if: 
\begin{equation*}
    \begin{split}
        \frac{\beta\sigma_{s}^{2}}{\sigma_{Q}^{2}+\beta\sigma_{s}^{2}}q+\frac{\sigma_{Q}^{2}}{\sigma_{Q}^{2}+\beta\sigma_{s}^{2}}s^{g}_{n3}-t(x_{n}-1)^{2}+\epsilon&>\frac{\sigma_{Q}^{2}}{\sigma_{Q}^{2}+\lambda'\beta\sigma_{s}^{2}}s^{g}_{n1}-tx_{n}^{2} 
        \text{  or, equivalently}\\
        \frac{\sigma_{Q}^{2}}{\sigma_{Q}^{2}+\beta\sigma_{s}^{2}}s^{g}_{n3}-\frac{\sigma_{Q}^{2}}{\sigma_{Q}^{2}+\lambda'\beta\sigma_{s}^{2}}s^{g}_{n1}&>t(1-2x_{n})-\epsilon-\frac{\beta\sigma_{s}^{2}}{\sigma_{Q}^{2}+\beta\sigma_{s}^{2}}q.
    \end{split}
\end{equation*}

For voters with $x_{n}<\frac{1}{2}$,  $\frac{\sigma_{Q}^{2}}{\sigma_{Q}^{2}+\beta\sigma_{s}^{2}}s^{g}_{n3}-\frac{\sigma_{Q}^{2}}{\sigma_{Q}^{2}+\lambda'\beta\sigma_{s}^{2}}s^{g}_{n1}\sim N\left(\frac{\sigma_{Q}^{2}}{\sigma_{Q}^{2}+\beta\sigma_{s}^{2}}q,\frac{2\sigma_{Q}^{4}}{\sigma_{Q}^{2}+\beta\sigma_{s}^{2}}\right)$. Therefore for a given $\epsilon$, the vote share of candidate 3 among the voters with $x_{n}<\frac{1}{2}$ equals to:
\begin{equation*}
    \int_{0}^{\frac{1}{2}}1-G\left((t(1-2x_{n})-\epsilon-q)\frac{\sqrt{\sigma^{2}_{Q}+\beta\sigma_{s}^{2}}}{\sqrt{2}\sigma_{Q}^{2}}\right)\,dx_{n},
\end{equation*}
where $G$ is the c.d.f.~of the standard normal distribution. On the other hand, for voters with $x_{n}>\frac{1}{2}$, $\frac{\sigma_{Q}^{2}}{\sigma_{Q}^{2}+\beta\sigma_{s}^{2}}s^{g}_{n3}-\frac{\sigma_{Q}^{2}}{\sigma_{Q}^{2}+\lambda'\beta\sigma_{s}^{2}}s^{g}_{n1}\sim N\left(\frac{\sigma_{Q}^{2}}{\sigma_{Q}^{2}+\sigma_{g}^{2}}q,\left(\frac{\sigma_{Q}^{4}}{\sigma_{Q}^{2}+\lambda\beta\sigma_{s}^{2}}+\frac{\sigma_{Q}^{4}}{\sigma_{Q}^{2}+\beta\sigma_{s}^{2}}\right)\right)$. For the given $\epsilon$, the vote received by candidate 3 from voters with $x_{n}>\frac{1}{2}$ equals:
\begin{equation*}
    \int^{1}_{\frac{1}{2}}1-G\left((t(1-2x_{n})-\epsilon-q)\left(\frac{\sigma_{Q}^{4}}{\sigma_{Q}^{2}+\lambda\beta\sigma_{s}^{2}}+\frac{\sigma_{Q}^{4}}{\sigma_{Q}^{2}+\beta\sigma_{s}^{2}}\right)^{-\frac{1}{2}}\right)\,dx_{n}.
\end{equation*}

Using the above distributions and vote shares, we can now characterize in Lemma~\ref{lem:onecand} the probability that the incumbent politician will be elected when there is only one challenger entering the primary. 

\medskip
\begin{lemma} \label{lem:onecand}
{\bf (Incumbent's chance of winning with one challenger)} When only one candidate on the left enters the race, the incumbent is elected with probability $G\left(\frac{-\epsilon_{13}^{*}}{\sigma_{\epsilon}}\right)$, where $\epsilon_{13}^{*}\leq -q$ is the solution of
{\small 
\begin{multline}
    \int_{0}^{\frac{1}{2}}G\left((t(1-2x_{n})-\epsilon_{13}^{*}-q)\frac{\sqrt{\sigma^{2}_{Q}+\beta\sigma_{s}^{2}}}{\sqrt{2}\sigma_{Q}^{2}}\right)\,dx_{n}+\\\int^{1}_{\frac{1}{2}}G\left((t(1-2x_{n})-\epsilon_{13}^{*}-q)\left(\frac{\sigma_{Q}^{4}}{\sigma_{Q}^{2}+\lambda\beta\sigma_{s}^{2}}+\frac{\sigma_{Q}^{4}}{\sigma_{Q}^{2}+\beta\sigma_{s}^{2}}\right)^{-\frac{1}{2}}\right)\,dx_{n}=\frac{1}{2}
    \label{eq:onecandidateepsilon}
\end{multline}
}
\end{lemma}
\medskip

Lemma~\ref{lem:onecand} shows that the incumbent is elected when the global taste shock of ideology favors him such that voters evaluate him highly, i.e., when $\epsilon$ is large. More specifically, the incumbent is elected when $\epsilon$ is larger
than the threshold~$\epsilon_{13}^{*}$ and with probability $1-G\left(\frac{\epsilon_{13}^{*}}{\sigma_{\epsilon}}\right)=G\left(\frac{-\epsilon_{13}^{*}}{\sigma_{\epsilon}}\right)$. The threshold $\epsilon_{13}^{*}$ thus measures the incumbent's advantage when only one challenger enters the race: the lower $\epsilon_{13}^{*}$ is, the larger is the incumbency advantage. We will use Lemma 1 along with the next lemmas in driving the results with regards to how the number of challengers entering the race influences the chances of the incumbent to get re-elected.

\paragraph{Two challengers enter the election.} Now suppose that both challengers, candidates $1$ and $2$, enter the election. We start by analyzing the outcome of the primary election and then move on to the analysis of the general election. The analysis resembles what we have carried out until now. Given the communication by politicians in the primary stage, $s^{p}_{n1}$ and $s^{p}_{n2}$, the expected valuation of the voters with $x_{n}<\frac{1}{2}$ of candidates $1$ and $2$ are
\begin{equation*}
    \begin{split}
        E(V(n,1)\mid s^{p}_{n1})&=\frac{\sigma_{Q}^{2}}{\sigma_{Q}^{2}+\sigma_{s}^{2}}s^{p}_{n1}-tx_{n}^{2}\\
        E(V(n,2)\mid s^{p}_{n2})&=\frac{\sigma_{Q}^{2}}{\sigma_{Q}^{2}+\sigma_{s}^{2}}s^{p}_{n2}-tx_{n}^{2}
    \end{split}
\end{equation*}
Since the signals received during the primary election for both candidates, $s^{p}_{n1}$ and $s^{p}_{n2}$, follow the same distribution ($N(0,\sigma_{Q}^{2}+\sigma_{p}^{2})$), candidates $1$ and $2$ each receive half the support from voters with $x_{n}<\frac{1}{2}$ during the primary stage and end up with a probability $\frac{1}{2}$ of winning the primary. Without loss of generality, let's assume  candidate $2$ wins the primary election and becomes the candidate for the general election on the left. 

To understand how a more competitive primary influences the outcome of a general election, let's first describe how these candidates are seen by the voters. 
During the general election period, voters receive additional information about candidates $2$ and $3$. Let voters with $x_{n}>\frac{1}{2}$ receive signals $s^{g}_{n2}$ and $s^{g}_{n3}$. In the following, as in the case with only one entrant, we characterize the vote share of the incumbent given $\epsilon$. As voters with $x_{n}>\frac{1}{2}$ do not receive additional information about candidate $2$ in the primary period, similar to the case with only one entrant, their expected valuation of candidates 2 and 3 becomes:  
\begin{equation}
    \begin{split}
        E(V(n,2)\mid s^{g}_{n2})&=\frac{\sigma_{Q}^{2}}{\sigma_{Q}^{2}+\lambda\beta\sigma_{s}^{2}}s^{g}_{n2}-tx_{n}^{2},\\
        E(V(n,3)\mid s^{g}_{n3})&=\frac{\beta\sigma_{s}^{2}}{\sigma_{Q}^{2}+\beta\sigma_{s}^{2}}q+\frac{\sigma_{Q}^{2}}{\sigma_{Q}^{2}+\beta\sigma_{s}^{2}}s^{g}_{n3}-t(x_{n}-1)^{2}+\epsilon.
    \end{split}
\end{equation}

Fixing $\epsilon$, the vote received by candidate $3$ from voters with right-wing views is the same as in the case with one challenger, i.e.,
\begin{equation*}
    \int^{1}_{\frac{1}{2}}1-G\left((t(1-2x_{n})-\epsilon-q)\left(\frac{\sigma_{Q}^{4}}{\sigma_{Q}^{2}+\lambda\beta\sigma_{s}^{2}}+\frac{\sigma_{Q}^{4}}{\sigma_{Q}^{2}+\beta\sigma_{s}^{2}}\right)^{-\frac{1}{2}}\right)\,dx_{n}.
\end{equation*}
Now, for voters with $x_{n}<\frac{1}{2}$, their expected evaluation towards candidate $2$ and $3$ given $s^{p}_{n2}$, $s^{g}_{n2}$ and $s^{g}_{n3}$ are
\begin{equation}
    \begin{split}
        E(V(n,2)\mid s^{p}_{n2},s^{g}_{n2})&=\frac{\frac{1}{\beta\sigma_{s}^{2}}}{\frac{1}{\sigma_{Q}^{2}}+\frac{1}{\beta\sigma_{s}^{2}}+\frac{1}{\sigma_{s}^{2}}}s^{g}_{n2}+\frac{\frac{1}{\sigma_{s}^{2}}}{\frac{1}{\sigma_{Q}^{2}}+\frac{1}{\beta\sigma_{s}^{2}}+\frac{1}{\sigma_{s}^{2}}}s^{p}_{n2}-tx_{n}^{2}\\
        E(V(n,3)\mid s^{g}_{n3})&=\frac{\beta\sigma_{s}^{2}}{\sigma_{Q}^{2}+\beta\sigma_{s}^{2}}q+\frac{\sigma_{Q}^{2}}{\sigma_{Q}^{2}+\beta\sigma_{s}^{2}}s^{g}_{n3}-t(x_{n}-1)^{2}+\epsilon.
    \end{split}
\end{equation}
Their evaluation of candidate $3$ follows the same expression as in the case where only one challenger enters the election. However, the expression of their evaluation of candidate $2$ is different as equation~\eqref{eq:evaluationonecandidate} because they have received information about candidate $2$ during the primary election. Fixing $\epsilon$, a voter with political ideology $x_{n}<\frac{1}{2}$ votes for candidate $3$ if and only if
\begin{equation*}
			\resizebox{1\hsize}{!}{$
    \begin{split}
        \frac{\beta\sigma_{s}^{2}}{\sigma_{Q}^{2}+\beta\sigma_{s}^{2}}q+\frac{\sigma_{Q}^{2}}{\sigma_{Q}^{2}+\beta\sigma_{s}^{2}}s^{g}_{n3}-t(x_{n}-1)^{2}+\epsilon&>\frac{\frac{1}{\beta\sigma_{s}^{2}}}{\frac{1}{\sigma_{Q}^{2}}+\frac{1}{\beta\sigma_{s}^{2}}+\frac{1}{\sigma_{s}^{2}}}s^{g}_{n2}+\frac{\frac{1}{\sigma_{s}^{2}}}{\frac{1}{\sigma_{Q}^{2}}+\frac{1}{\beta\sigma_{s}^{2}}+\frac{1}{\sigma_{s}^{2}}}s^{p}_{n2}-tx_{n}^{2}\\
        \frac{\sigma_{Q}^{2}}{\sigma_{Q}^{2}+\beta\sigma_{s}^{2}}s^{g}_{n3}-\frac{\frac{1}{\beta\sigma_{s}^{2}}}{\frac{1}{\sigma_{Q}^{2}}+\frac{1}{\beta\sigma_{s}^{2}}-\frac{1}{\sigma_{s}^{2}}}s^{g}_{n2}+\frac{\frac{1}{\sigma_{s}^{2}}}{\frac{1}{\sigma_{Q}^{2}}+\frac{1}{\beta\sigma_{s}^{2}}+\frac{1}{\sigma_{s}^{2}}}s^{p}_{n2}&>t(1-2x_{n})-\epsilon-\frac{\beta\sigma_{s}^{2}}{\sigma_{Q}^{2}+\beta\sigma_{s}^{2}}q.
    \end{split}
    $}
\end{equation*}
Note that, 
\begin{multline*}
    \frac{\sigma_{Q}^{2}}{\sigma_{Q}^{2}+\beta\sigma_{s}^{2}}s^{g}_{n3}-\frac{\frac{1}{\beta\sigma_{s}^{2}}}{\frac{1}{\sigma_{Q}^{2}}+\frac{1}{\beta\sigma_{s}^{2}}+\frac{1}{\sigma_{s}^{2}}}s^{g}_{n2}-\frac{\frac{1}{\sigma_{s}^{2}}}{\frac{1}{\sigma_{Q}^{2}}+\frac{1}{\beta\sigma_{s}^{2}}+\frac{1}{\sigma_{s}^{2}}}s^{p}_{n2}\sim\\ N\left(\frac{\sigma_{Q}^{2}}{\sigma_{Q}^{2}+\beta\sigma_{s}^{2}}q,\frac{\sigma_{Q}^{4}}{\sigma_{Q}^{2}+\beta\sigma_{s}^{2}}+\frac{\frac{\sigma_{Q}^{2}+\beta\sigma_{s}^{2}}{\beta^{2}\sigma_{s}^{4}}}{(\frac{1}{\sigma_{Q}^{2}}+\frac{1}{\beta\sigma_{s}^{2}}+\frac{1}{\sigma_{s}^{2}})^{2}}+\frac{\frac{\sigma_{Q}^{2}+\sigma_{s}^{2}}{\sigma_{s}^{4}}}{(\frac{1}{\sigma_{Q}^{2}}+\frac{1}{\beta\sigma_{s}^{2}}+\frac{1}{\sigma_{s}^{2}})^{2}}\right).
\end{multline*}
Thus, the vote received by candidate $3$ from voters with $x_{n}<\frac{1}{2}$ is equal to
\begin{equation*}
			\resizebox{1\hsize}{!}{$
    \int_{0}^{\frac{1}{2}}1-G\left((t(1-2x_{n})-\epsilon-q)\left(\frac{\sigma_{Q}^{4}}{\sigma_{Q}^{2}+\beta\sigma_{s}^{2}}+\frac{\frac{\sigma_{Q}^{2}+\beta\sigma_{s}^{2}}{\beta^{2}\sigma_{s}^{4}}}{(\frac{1}{\sigma_{Q}^{2}}+\frac{1}{\beta\sigma_{s}^{2}}+\frac{1}{\sigma_{s}^{2}})^{2}}+\frac{\frac{\sigma_{Q}^{2}+\sigma_{s}^{2}}{\sigma_{s}^{4}}}{(\frac{1}{\sigma_{Q}^{2}}+\frac{1}{\beta\sigma_{s}^{2}}+\frac{1}{\sigma_{s}^{2}})^{2}}\right)^{-\frac{1}{2}}\right)\,dx_{n}.
    $}
\end{equation*}
Similar to Lemma~\ref{lem:onecand}, the vote received by candidate $3$ is increasing in $\epsilon$, thus he is elected if and only if $\epsilon$ is bigger than some threshold which gives us the following result.

\begin{lemma}\label{lem:twocand}
When two challengers from the left enter the race, the incumbent is elected with probability $G\left(\frac{-\epsilon_{123}^{*}}{\sigma_{\epsilon}}\right)$, where $\epsilon_{123}^{*}$ is the solution to
\footnotesize
\begin{multline}
   \int_{0}^{\frac{1}{2}}G\left((t(1-2x_{n})-\epsilon_{123}^{*}-q)\left(\frac{\sigma_{Q}^{4}}{\sigma_{Q}^{2}+\beta\sigma_{s}^{2}}+\frac{\frac{\sigma_{Q}^{2}+\beta\sigma_{s}^{2}}{\beta^{2}\sigma_{s}^{4}}}{(\frac{1}{\sigma_{Q}^{2}}+\frac{1}{\beta\sigma_{s}^{2}}+\frac{1}{\sigma_{s}^{2}})^{2}}+\frac{\frac{\sigma_{Q}^{2}+\sigma_{s}^{2}}{\sigma_{s}^{4}}}{(\frac{1}{\sigma_{Q}^{2}}+\frac{1}{\beta\sigma_{s}^{2}}+\frac{1}{\sigma_{s}^{2}})^{2}}\right)^{-\frac{1}{2}}\right)\,dx_{n}\\+\int^{1}_{\frac{1}{2}}G\left((t(1-2x_{n})-\epsilon_{123}^{*}-q)\left(\frac{\sigma_{Q}^{4}}{\sigma_{Q}^{2}+\lambda\beta\sigma_{s}^{2}}+\frac{\sigma_{Q}^{4}}{\sigma_{Q}^{2}+\beta\sigma_{s}^{2}}\right)^{-\frac{1}{2}}\right)\,dx_{n}=\frac{1}{2}. 
\label{eq:twocandidateepsilon}
\end{multline}
\normalsize
\end{lemma}

Similar to Lemma~\ref{lem:onecand}, the threshold $\epsilon_{123}^{*}$ in Lemma~\ref{lem:twocand} measures the incumbency advantage when two challengers enter the race: the lower $\epsilon_{123}^{*}$ is, the greater is the incumbency advantage and the incumbent is elected with a higher probability.
Next, using this lemma, we can compare the winning probability of the incumbent with one or two challengers on the left and show that a higher number of challengers, or {\em more} competition, could benefit the incumbent. 

Note that, as the left hand side of Equation~\eqref{eq:twocandidateepsilon} is decreasing in $\epsilon_{123}^{*}$, candidate $3$ wins with a higher probability when both challengers on the left enter, i.e., $\epsilon_{123}^{*}<\epsilon_{13}^{*}$, iff
\footnotesize
\begin{multline}
    \int_{0}^{\frac{1}{2}}G\left((t(1-2x_{n})-\epsilon_{13}^{*}-q)\left(\frac{\sigma_{Q}^{4}}{\sigma_{Q}^{2}+\beta\sigma_{s}^{2}}+\frac{\frac{\sigma_{Q}^{2}+\beta\sigma_{s}^{2}}{\beta^{2}\sigma_{s}^{4}}}{(\frac{1}{\sigma_{Q}^{2}}+\frac{1}{\beta\sigma_{s}^{2}}+\frac{1}{\sigma_{s}^{2}})^{2}}+\frac{\frac{\sigma_{Q}^{2}+\sigma_{s}^{2}}{\sigma_{s}^{4}}}{(\frac{1}{\sigma_{Q}^{2}}+\frac{1}{\beta\sigma_{s}^{2}}+\frac{1}{\sigma_{s}^{2}})^{2}}\right)^{-\frac{1}{2}}\right)\,dx_{n}<\\\int_{0}^{\frac{1}{2}}G\left((t(1-2x_{n})-\epsilon_{13}^{*}-q)\frac{\sqrt{\sigma^{2}_{Q}+\beta\sigma^{2}_{s}}}{\sqrt{2}\sigma_{Q}^{2}}\right)\,dx_{n}.\label{eq:twooronecandidates}
\end{multline}
\normalsize

The incumbent candidate $3$ is elected with a higher probability if he gets more votes from voters with $x_{n}<\frac{1}{2}$ given $\epsilon=\epsilon_{13}^{*}$. Now note that by Lemma~\ref{lem:onecand}, $\epsilon_{13}^{*}\leq -q$ and thus $t(1-2x_{n})-\epsilon_{13}^{*}-q>0$ for all $x_{n}\in[0,\frac{1}{2}]$. It implies that without any information, all voters with $x_{n}<\frac{1}{2}$ would vote for the challenger candidate $2$. Thus, inequality~\eqref{eq:twooronecandidates} is equivalent to
\begin{equation}
    \frac{\frac{\sigma_{Q}^{2}+\beta\sigma_{s}^{2}}{\beta^{2}\sigma_{s}^{4}}}{(\frac{1}{\sigma_{Q}^{2}}+\frac{1}{\beta\sigma_{s}^{2}}+\frac{1}{\sigma_{s}^{2}})^{2}}+\frac{\frac{\sigma_{Q}^{2}+\sigma_{s}^{2}}{\sigma_{s}^{4}}}{(\frac{1}{\sigma_{Q}^{2}}+\frac{1}{\beta\sigma_{s}^{2}}+\frac{1}{\sigma_{s}^{2}})^{2}}>\frac{\sigma_{Q}^{4}}{\sigma_{Q}^{2}+\beta\sigma_{s}^{2}}.
    \label{eq:twooronecandidates1}
\end{equation}
Put differently, incumbent candidate $3$ is elected with a higher probability when both, instead of only one, challengers enter the race if and only if the evaluation of the challenger candidate $2$ from voters with left-wing views, i.e., $x_{n}<\frac{1}{2}$, is noisier. Proposition~\ref{prop:morechallengers} demonstrates this key finding. 

\medskip
\begin{proposition}{\bf (Number of Challengers and Incumbency Advantage)}
Candidate $3$, or the incumbent, wins the general election with a higher probability when there are two challengers instead of one if $\frac{\sigma^{2}_{s}}{\sigma^{2}_{Q}}$ or $\beta$ is sufficiently high. Formally, $\epsilon_{123}^{*}<\epsilon_{13}^{*}$ holds iff:
\begin{equation}
    \frac{\sigma^{2}_{s}}{\sigma^{2}_{Q}}>\frac{\sqrt{(\beta-1)^{2}+8}-(\beta-1)}{2\beta}.
    \label{eq:propmorechallengers}
\end{equation}
\label{prop:morechallengers}
\end{proposition}

Proposition~\ref{prop:morechallengers} shows that, counter-intuitively, the incumbent could win with a higher probability when there are more challengers entering an electoral race. In this case, {\em more} competition strengthens the incumbency advantage. To see the intuition for this result, consider how a signal received in the primary stage influences the likelihood of a challenger winning in the general election, depending on the number of challengers entering the race.  
In terms of purely their ideological match, the voters with $x_{n}<\frac{1}{2}$ prefer the challenger candidates over the incumbent before receiving any signals. However, importantly, these voters are targeted during the primaries with additional information received through various media. When some voters draw negative signals about a challenger candidate, some may update their valuation of a candidate downward, and those at the ideological margin may switch to vote for the incumbent in the general election. The noisier the evaluation of the challenger is, the higher is the probability that they will get a negative signal and switch to the incumbent. When $\frac{\sigma_{s}^{2}}{\sigma_{Q}^{2}}$ is high, the extra information that left-wing voters receive in the primary is noisier, which makes their evaluation of the challengers noisier. As a result, there is a higher probability that voters with $x_n <\frac{1}{2}$ update their beliefs negatively after drawing negative signals in the primary, and this benefits the incumbent. The incumbent then wins with a higher probability if the evaluation towards the challengers from the constituency they target ($x_n <\frac{1}{2}$) is noisier. 

When $\beta$ is high, the information received in the primary period is more informative than that in the general election and therefore voters put more weight on them in their belief updating. A negative signal received during the primary stage thus has a heavier weight in the voter's evaluation. On the other hand, if they receive a positive signal in the primary stage, although it protects the challenger from negative signals in the general election, the incremental benefit is small, because the information in the general election weighs less in voters' evaluation and is unlikely to affect voters' decision. Thus, there is an asymmetric effect of good news and bad news in the primary election for the incumbent and the challenger. 

This result also highlights the potential adverse effects of noisier political communication arriving at later stages of an election. Such noise may arrive from misinformation campaigns or advertisements with false information launched closer to the election date. Such efforts are more likely to hurt the challengers rather than the incumbents. Similarly, political communication restrictions brought on at later stages of elections to reduce political information may strengthen an incumbent's chances of winning. In a related recent development, Facebook announced that it will ban political advertisements during the one week period before the 2020 U.S. Presidential election. This restriction, which comes late in the election period, is an example of a policy which reduces information received during the general election stage compared to the primary stage, and based on the predictions from our model, may disproportionately harm the challengers rather than the incumbent politician.
%


\subsection{Equilibrium Characterization}

Now we are ready to characterize the subgame perfect Nash equilibrium. We focus on pure strategy equilibria.\footnote{In the mixed strategy equilibria, there might be a coordination failure among candidates on the left, in which case no candidates on the left enter the election with strictly positive probability. This type of equilibria is unrealistic and is therefore ignored.} We also assume a candidate chooses to enter the race in case of indifference.
Proposition~\ref{prop:lowerC} investigates how the changes in the fixed cost of entering a market alters the competitiveness of political races and the resulting probability of the incumbent's chances of winning in an election. Such fixed costs may be associated with the initial fundraising, procedural challenges, or the difficulty of initiating marketing and political communication. 

\begin{proposition}{\bf (Cost of Entry and Competitiveness of Races)}
When $\frac{1}{2}G\left(\frac{\epsilon_{123}^{*}}{\sigma_{\epsilon}}\right)<G\left(\frac{\epsilon_{13}^{*}}{\sigma_{\epsilon}}\right)$, the equilibrium is characterized as follows:
\begin{enumerate}
    \item When the cost of entering a political race is small ($C\leq\frac{1}{2}G\left(\frac{\epsilon_{123}^{*}}{\sigma_{\epsilon}}\right)$), both candidates $1$ and $2$ enter the election and win with probability $\frac{1}{2}G\left(\frac{\epsilon_{123}^{*}}{\sigma_{\epsilon}}\right)$ and the incumbent (candidate $3$) wins with probability $G\left(\frac{-\epsilon_{123}^{*}}{\sigma_{\epsilon}}\right)$;
    \item when entry has an intermediate cost ($C\in\Bigl(\frac{1}{2}G\left(\frac{\epsilon_{123}^{*}}{\sigma_{\epsilon}}\right),G\left(\frac{\epsilon_{13}^{*}}{\sigma_{\epsilon}}\right)\Bigr]$), only one of candidates $1$ and $2$ enters the race and win with probability $\frac{1}{2}G\left(\frac{\epsilon_{13}^{*}}{\sigma_{\epsilon}}\right)$, while candidate $3$ wins with probability $G\left(\frac{-\epsilon_{13}^{*}}{\sigma_{\epsilon}}\right)$;
    \item when the entry cost is high ($C>G\left(\frac{\epsilon_{13}^{*}}{\sigma_{\epsilon}}\right)$), neither of candidates $1$ and $2$ enters the race and the incumbent wins with probability $1$.
\end{enumerate}
Otherwise, when $\frac{1}{2}G\left(\frac{\epsilon_{123}^{*}}{\sigma_{\epsilon}}\right)\geq G\left(\frac{\epsilon_{13}^{*}}{\sigma_{\epsilon}}\right)$, the equilibrium is characterized as follows:
\begin{enumerate}
    \item when the cost of entering a race is small ($C\leq\frac{1}{2}G\left(\frac{\epsilon_{123}^{*}}{\sigma_{\epsilon}}\right)$), both candidates $1$ and $2$ enter the election and win with probability $\frac{1}{2}G\left(\frac{\epsilon_{123}^{*}}{\sigma_{\epsilon}}\right)$ while the incumbent wins with probability $G\left(\frac{-\epsilon_{123}^{*}}{\sigma_{\epsilon}}\right)$;
    \item when the cost is high ($C>\frac{1}{2}G\left(\frac{\epsilon_{123}^{*}}{\sigma_{\epsilon}}\right)$), candidates $1$ and $2$ both do not enter the election and the incumbent wins with probability $1$.
\end{enumerate}
\label{prop:lowerC}
\end{proposition}

Proposition indicates the key and intuitive finding that, a high cost of entering politics sets a barrier to entering politics, therefore reducing the competitiveness of races. A lower $C$, for instance due to the availability of social media and cheaper digital advertising,  incentivizes more challengers to enter the election and thus promotes competition. This finding is followed by the counter-intuitive result that an increase in competition due to higher number of challengers could, in fact, benefit the incumbent by increasing his chances of winning an election in Corollary~\ref{coro:lowercbenefitincumbent}.

\medskip
\begin{corollary}{\bf (Cost of Entry and Incumbency Advantage)}
As the cost of entering a race ($C$) decreases, incumbency advantage, or candidate $3$'s probability of winning the race is higher if $\frac{1}{2}G\left(\frac{\epsilon_{123}^{*}}{\sigma_{\epsilon}}\right)<G\left(\frac{\epsilon_{13}^{*}}{\sigma_{\epsilon}}\right)$ and $C<G\left(\frac{\epsilon_{13}^{*}}{\sigma_{\epsilon}}\right)$, and $\frac{1}{2}G\left(\frac{\epsilon_{123}^{*}}{\sigma_{\epsilon}}\right)<G\left(\frac{\epsilon_{13}^{*}}{\sigma_{\epsilon}}\right)$ holds if:
\begin{equation*}
    \frac{\sigma^{2}_{s}}{\sigma^{2}_{Q}}>\frac{\sqrt{(\beta-1)^{2}+8}-(\beta-1)}{2\beta}.
\end{equation*}
\label{coro:lowercbenefitincumbent}
\end{corollary}

Corollary~\ref{coro:lowercbenefitincumbent} demonstrates that the incumbency advantage strengthens with reducing the barriers to entering electoral races when information throughout the election is noisy, or when the information in the primary weighs more relative to the information in the general election. When Equation~\eqref{eq:propmorechallengers} holds, a decrease in the cost of entry to a race ($C$), which increases the number of challengers from $1$ to $2$, increases the probability of the incumbent's re-election. Proposition~\ref{prop:lowerC} and Corollary~\ref{coro:lowercbenefitincumbent} deliver the main message of the paper: although a lower entry cost induces more competition, i.e., more challengers to enter the race, this does not necessarily weaken incumbency advantage. In fact, reducing barriers to entry could benefit the incumbent and boost his probability of being re-elected. 

While our main findings are in the context of cost of marketing as an entry barrier to politics, parallel arguments can be brought up if new and similarly positioned firms first face a competition among themselves. Such a competition may be for startup funding, regional distribution competition, or competition for shelf-space of a retailer before facing a nationally known and well-advertised incumbent. Our analysis suggests that, availability of social media and cheap digital advertising may facilitate entry of new brands which are differentiated from the incumbent, but intensified competition may not be sufficient to weaken the market share of the incumbent. 

In the following, we analyze how the perceived quality of the incumbent and other parameters about the informativeness of signals in primary and general election affect the incumbency advantage.

\medskip
\begin{proposition}\label{prop:qandlambda}
{\bf (Perceived Quality and Informational Advantage of Incumbent)}

\noindent {\bf \em (i)}  Fixing the number of challengers, the incumbent is re-elected with a higher probability if he has a higher perceived quality or a higher informational advantage, i.e., $\epsilon_{13}^{*}$ and $\epsilon_{123}^{*}$ both increase in $q$ and $\lambda$.

\noindent {\bf \em (ii)} When 
more competition reduces incumbency advantage, then the incumbent is re-elected with a higher probability when $q$ and $\lambda$ increase.
In contrast, when 
more competition strengthens incumbency advantage, the effects of $q$ and $\lambda$ are non-monotonic. 

\end{proposition}
\medskip

The proposition demonstrates how the quality advantage (i.e., the expected match value $q$) and the informational advantage ($\lambda$) of the incumbent impact the re-election probability of the incumbent. Part (i) indicates that a higher expected match value of the incumbent ($q$) implies a higher winning probability of an incumbent. As a candidate increases in his appeal to voters, naturally, he gains more support.  A higher informational advantage relative to the challengers, for instance, the recent bans on political advertisements which limit political outreach to voters, is similarly likely to hurt the challenger's ability to convince the voters to switch and vote for him and increase the probability that the incumbent will win the general election. The effect of an increase in the two variables, when the number of entrants to the race is exogenous, is straightforward and is hurtful to the chances of a newcomer. Part (ii) of the proposition demonstrates how, when the number of entrants to the race is also endogenously determined by the changes to these parameters, the incumbency advantage is altered. Since an increase in either parameter discourages entry of challengers, 
when more competition reduces the incumbency advantage, a higher $q$ or $\lambda$ implies reduced number of challengers entering a race, and an increase in the incumbency advantage for two reasons: first, there are fewer number of challengers entering the race, and second the challengers who enter are more limited in competing against the incumbent. 
If more competition strengthens the incumbency advantage, however, the first effect reverses. A higher $q$ or $\lambda$ implies fewer number of challengers entering a race and a decrease in the incumbency advantage while still indicating a more limited ability of the challengers competing against the incumbent, resulting in an ambiguous outcome in terms of the change in incumbency advantage. The key point is that, in both scenarios, when political advertisement bans of online platforms reduce the ability of the challengers to disseminate information about their candidacy more than they reduce that of the incumbent, this may alter the results of elections in a way to favor known, career-politicians.


\subsection{Generalizing Results to More than Two Challengers}\label{sec:morethan2challengers}

We now generalize the key finding that more competition could strengthen incumbency advantage to settings with more than two challengers. We capture the increase in competition in the primary through its effect on the information structure in this stage. More specifically, we assume that more challengers lead to a more informative signal structure in the primary as they have to fight harder for attention, for which we provide a micro-foundation in Appendix B. In the appendix, we show without unnecessarily complicating the derivations that, the challengers have more incentives to engage in communication (e.g., more media coverage, debates, discussions or activities on social media) in a more competitive primary. 
Formally, let's denote the number of challenger candidates as $e$. We generalize the baseline model such that the parameters of the signal structures $\sigma_{s}^{2}(e)$ and $\beta(e)$ are functions of $e$. Based on our discussion above, more challengers entering an electoral race induces more communication during the primary stage, but does not affect the communication in the general election. To capture this relationship, we assume that the precision of the signal in the primary stage increases in $e$, i.e., $\sigma_{s}^{2}(e)$ decreases in $e$, while the precision of signals in the general stage remains a constant. That is, $\beta(e)\sigma_{s}^{2}(e)$ is invariant in $e$. In this scenario, we have the following result.


\medskip
\begin{corollary} {\bf (More Competition in the Primary and Incumbency Advantage)}
Suppose the competition in primary period intensifies such that the signals obtained in the primary are more precise relative to the information provision in the general election. 
Suppose that the number of challengers increases from $e$ to $e+1$, the incumbent wins with a higher probability if
\begin{equation*}
    \frac{\beta(e)\sigma^{2}_{s}(e)}{\sigma^{2}_{Q}}>\frac{\sqrt{(\beta(e)-1)^{2}+8}-(\beta(e)-1)}{2}.
\end{equation*}
In words, the incumbency advantage is stronger if either $\sigma_{s}^{2}(e)$ is sufficiently small  or if $\beta(e)$ is sufficiently large. Moreover, if the incumbency advantage is stronger when the number of challengers increases from $e$ to $e+1$, it is also stronger when the number of challengers increases from $e+1$ to $e+2$. 
\label{coro:morechallengers}
\end{corollary}
\medskip

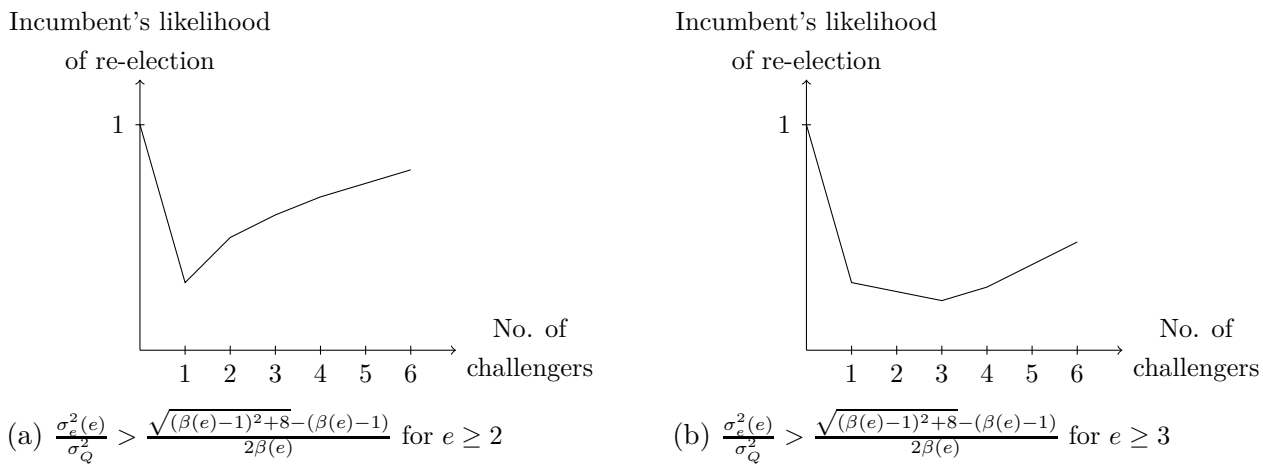
\begin{figure}[t]
    \centering
    \begin{subfigure}[b]{0.4\linewidth}
    \begin{tikzpicture}[scale=0.6]
    \draw[->] (0,0)--(7,0) node[right, align=center] {\footnotesize No. of\\\footnotesize challengers};
    \draw (1,0.1)--(1,-0.1) node[below] {\footnotesize $1$};
    \draw (2,0.1)--(2,-0.1) node[below] {\footnotesize $2$};
    \draw (3,0.1)--(3,-0.1) node[below] {\footnotesize $3$};
    \draw (4,0.1)--(4,-0.1) node[below] {\footnotesize $4$};
    \draw (5,0.1)--(5,-0.1) node[below] {\footnotesize $5$};
    \draw (6,0.1)--(6,-0.1) node[below] {\footnotesize $6$};
    \draw (0.1,5)--(-0.1,5) node[left] {\footnotesize $1$};
    \draw[->] (0,0)--(0,6) node[above, align=center] {\footnotesize Incumbent's likelihood \\\footnotesize of re-election};
    \draw (0,5)--(1,1.5)--(2,2.5)--(3,3)--(4,3.4)--(5,3.7)--(6,4);
    \end{tikzpicture}
    \caption{\footnotesize $\frac{\sigma^{2}_{e}(e)}{\sigma^{2}_{Q}}>\frac{\sqrt{(\beta(e)-1)^{2}+8}-(\beta(e)-1)}{2\beta(e)}$ for $e\geq 2$}
        \label{fig:numberofchallengers1}
    \end{subfigure}
     \hspace{0.7in}
     \begin{subfigure}[b]{0.4\linewidth}    
     \begin{tikzpicture}[scale=0.6]
    \draw[->] (0,0)--(7,0) node[right, align=center] {\footnotesize No. of\\\footnotesize challengers};
    \draw (1,0.1)--(1,-0.1) node[below] {\footnotesize $1$};
    \draw (2,0.1)--(2,-0.1) node[below] {\footnotesize $2$};
    \draw (3,0.1)--(3,-0.1) node[below] {\footnotesize $3$};
    \draw (4,0.1)--(4,-0.1) node[below] {\footnotesize $4$};
    \draw (5,0.1)--(5,-0.1) node[below] {\footnotesize $5$};
    \draw (6,0.1)--(6,-0.1) node[below] {\footnotesize $6$};
    \draw (0.1,5)--(-0.1,5) node[left] {\footnotesize $1$};
    \draw[->] (0,0)--(0,6) node[above, align=center] {\footnotesize  Incumbent's likelihood  \\\footnotesize of re-election};
    \draw (0,5)--(1,1.5)--(2,1.3)--(3,1.1)--(4,1.4)--(5,1.9)--(6,2.4);
    \end{tikzpicture}
    \caption{\footnotesize $\frac{\sigma^{2}_{e}(e)}{\sigma^{2}_{Q}}>\frac{\sqrt{(\beta(e)-1)^{2}+8}-(\beta(e)-1)}{2\beta(e)}$ for $e\geq 3$}
        \label{fig:numberofchallengers2}
    \end{subfigure}
    \caption{\footnotesize Illustration of how the number of challengers affects the incumbent's probability of being re-elected. } 
    \label{fig:numberofchallengers}
\end{figure}

The results from the corollary are illustrated in Figure~\ref{fig:numberofchallengers}. In Figure~\ref{fig:numberofchallengers1}, when $\frac{\sigma^{2}_{s}(e)}{\sigma^{2}_{Q}}>\frac{\sqrt{(\beta(e)-1)^{2}+8}-(\beta(e)-1)}{2\beta(e)}$ for all $e\geq 2$ holds, incumbent's probability of re-election increases in the number of challengers with two or more candidates. 
When the number of challengers increases from $0$ to $1$, as expected, the incumbent's chance of being re-elected drops as he now faces competition. In (a), the likelihood increases with more entrants beyond a single challenger, in line with Corollary~\ref{coro:morechallengers} when Equation~\eqref{eq:propmorechallengers} holds for $e\geq 2$. The entry of each additional challenger strengthens the incumbent's chance of re-election. In Figure~\ref{fig:numberofchallengers2}, when $\frac{\sigma^{2}_{s}(e)}{\sigma^{2}_{Q}}\leq\frac{\sqrt{(\beta(e)-1)^{2}+8}-(\beta(e)-1)}{2\beta(e)}$ for $e\geq 3$ holds, the probability of re-election first decreases and then increases in the number of challengers with three or more challengers, in line with  Corollary~\ref{coro:morechallengers} where Equation~\eqref{eq:propmorechallengers} holds for a sufficiently large $e$, i.e., in this example when $e\geq 3$. 

The intuition of the result is very similar to Proposition~\ref{prop:morechallengers}. During the primary period, the challengers are interested primarily in communicating with the voters who vote in the primary: the voters whose ideology is aligned with theirs. When there is more competition in the primary, there is more communication with the voters and it induces a more precise signal in the primary relative to the general election stage ($\sigma_{s}^{2}(e)$ decreases and $\beta(e)$ increases). As a result, the information from the primary weighs more on the evaluation of voters with left-wing views. Negative information has a larger adverse effect on the challenger's chances of being elected. On the other hand, although a positive signal in the primary protects the challenger from negative news in the general election, this positive effect is small as the information in the general election weighs less in voters' evaluation and thus is unlikely to persuade left-wing voters to switch. In this case, the negative news in the primary hurts the challengers more than the positive news benefit them, and as a result more competition strengthens the incumbency advantage.


\section{Conclusion} \label{sec: conclusion}

Incumbency advantage has grown steadily since the 1940ies \citep{Ansolabehere00}. Candidates who have been elected to an office once hold continuing advantages over their opponents, which remains an important barrier to making elections more competitive. A key source of this advantage is the difference between the incumbents and challengers in their ability to run marketing and communication campaigns. More specifically, the difference in ability to access media to inform and persuade voters, either because experienced politicians are more likely to be covered in media or because newcomers lack funding to buy advertising or other forms of paid messaging, resulted in persistent re-election success of incumbents. 
 
Internet, digital advertising, and social media relaxed this limitation by giving new politicians a platform to communicate with masses \citep{petrova2020social} thereby reducing informational barriers to enter politics. Many political newcomers now communicate with their constituency via social media such as Facebook and Twitter to inform and persuade them, which makes electoral races more competitive. But does more competition necessarily help to reverse incumbency advantage? 
We answer this question considering a specific informational environment: marketing campaigns in political races. We develop a model where incumbency advantage can come into play through two channels. First, voters may hold a positive prior about the match value of the incumbent. Second, the incumbent may hold structural advantages in reaching out to voters who are ideologically different than their base relative to  challengers. 

We find, first, that lowering cost of communication via digital advertising and social media reduces the barriers to entering politics and will make races more competitive. But higher number of challengers, and resulting higher levels of marketing and communication campaigns during the primary, do not necessarily mitigate incumbency advantage, and may in fact strengthen it. The incumbent can benefit from intensified marketing efforts of other candidates during the primary.  
Specifically, compared to an election where there are fewer challengers, an election with more than two challengers may increase an incumbent's chances of re-election. This is because of the asymmetric effect of negative versus positive information during the primary that voters use to resolve uncertainty about the candidates. Challengers' communication during the primary targets the individuals who vote in the primaries, and these are the individuals whose political opinions are aligned with that of the  challengers. Those voters ex-ante prefer challengers, but upon receiving negative information in the primary, they might switch to vote for the incumbent. Thus positive information at the primary stage is unlikely to gain a positional advantage for the challengers. Negative news is more likely to dominate the impact of positive news, especially when communication during the primary is more informative than the general election.

Second, we find that restricting political advertising and micro-targeting could hurt the challengers disproportionately more than the incumbent, when incumbents hold strategic advantages in accessing the entire body of voters. This finding implies that recent political advertising and micro-targeting bans instituted by online platforms such as Twitter, Facebook, and Google may ultimately hurt the chances of challengers in electoral races.  Managers of online social media platforms should be cognizant of the decisions they make regarding advertising policies, as these policies will influence the outcome of elections, despite the intention to protect consumers.

Our findings demonstrate the key benefits and costs of cheaper communications technology facilitated by digital advertising and social media platforms. This topic is very timely and important, and our study is focusing on the outcome of these changes on political competition. There may be other effects of easily accessing such platforms which are peripheral to our study. Future research can consider other factors. Researchers can also empirically test the predictions of this study to test how incumbents fare in elections compared to newcomers, as tools of communication got cheaper over the years or as political advertising bans have been erected by online platforms, using a large sample size of politicians.

\setstretch{1.1}
{\small 
\bibliographystyle{chicago}
\bibliography{lit}

\vspace{0.4in}

\small{

\begin{center}
    {\bf FOR ONLINE PUBLICATION ONLY }
\end{center}

\subsection*{\large Appendix A: {\small Proofs of Propositions, Lemmas, and Corollaries}} \label{proof}


\noindent{\bf Proof of Lemma~\ref{lem:onecand}: }
Here we show that $\epsilon_{13}^{*}\leq -q$. As will be shown in Proposition~\ref{prop:qandlambda}, $\epsilon_{13}^{*}$ decreases in $\lambda$. Moreover, note that when $\lambda=1$, $\epsilon_{13}^{*}=-q$ satisfies Equation~\ref{eq:onecandidateepsilon}. Thus as $\lambda\geq 1$, $\epsilon_{13}^{*}\leq -q$.

\medskip

\noindent {\bf{Proof of Proposition~\ref{prop:morechallengers}:}}
    Inequality~\ref{eq:twooronecandidates1} can be rewritten as follows:
    \begin{equation*}
        \begin{split}
             \frac{\frac{\sigma_{Q}^{2}+\beta\sigma_{s}^{2}}{\beta^{2}\sigma_{s}^{4}}}{(\frac{1}{\sigma_{Q}^{2}}+\frac{1}{\beta\sigma_{s}^{2}}+\frac{1}{\sigma_{s}^{2}})^{2}}+\frac{\frac{\sigma_{Q}^{2}+\sigma_{s}^{2}}{\sigma_{s}^{4}}}{(\frac{1}{\sigma_{Q}^{2}}+\frac{1}{\beta\sigma_{s}^{2}}+\frac{1}{\sigma_{s}^{2}})^{2}}&>\frac{\sigma_{Q}^{4}}{\sigma_{Q}^{2}+\beta\sigma_{s}^{2}}\\
             \frac{\sigma_{Q}^{4}\sigma_{s}^{4}(\sigma_{Q}^{2}+\beta\sigma_{s}^{2})+\sigma_{Q}^{4}\beta^{2}\sigma_{s}^{4}(\sigma_{Q}^{2}+\sigma_{s}^{2})}{(\sigma_{Q}^{2}\beta\sigma_{s}^{2}+\sigma_{Q}^{2}\sigma_{s}^{2}+\beta\sigma_{s}^{4})^{2}}&>\frac{\sigma_{Q}^{4}}{\sigma_{Q}^{2}+\beta\sigma_{s}^{2}}\\
             \frac{(\beta^{2}+1)\sigma_{Q}^{2}+(\beta^{2}+\beta)\sigma_{s}^{2}}{(\sigma_{Q}^{2}(\beta+1)+\beta\sigma_{s}^{2})^{2}}&>\frac{1}{\sigma_{Q}^{2}+\beta\sigma_{s}^{2}}\\
             (\beta^{2}+1)\sigma_{Q}^{4}+(\beta^{2}+1)\beta\sigma_{Q}^{2}\sigma_{s}^{2}+(\beta^{2}+\beta)\sigma_{s}^{2}\sigma_{Q}^{2}+(\beta^{2}+\beta)\beta\sigma_{s}^{4}&>(\sigma_{Q}^{2}(\beta+1)+\beta\sigma_{s}^{2})^{2}\\
             \beta^{2}(\beta-1)\sigma_{Q}^{2}\sigma_{s}^{2}+\beta^{3}\sigma_{s}^{4}&>2\beta\sigma_{Q}^{4}\\
             \beta^{2}\frac{\sigma^{4}_{s}}{\sigma^{4}_{Q}}+\beta(\beta-1)\frac{\sigma^{2}_{s}}{\sigma^{2}_{Q}}&>2\\
             \left(\frac{\sigma^{2}_{s}}{\sigma^{2}_{Q}}+\frac{\beta-1}{2\beta}\right)^{2}&>\frac{8+(\beta-1)^{2}}{4\beta^{2}}\\
             \frac{\sigma^{2}_{s}}{\sigma^{2}_{Q}}&>\frac{\sqrt{(\beta-1)^{2}+8}-(\beta-1)}{2\beta}
        \end{split}
    \end{equation*}
    The second part of the proposition follows from:
    \begin{equation*}
        \begin{split}
            &\frac{\partial}{\partial\beta}\left(\frac{\sqrt{(\beta-1)^{2}+8}-(\beta-1)}{2\beta}\right)\\
            =&\frac{1}{2\beta^{2}}\left(\frac{\beta}{2\sqrt{(\beta-1)^{2}+8}}-\beta-(\sqrt{(\beta-1)^{2}+8}-(\beta-1))\right)\\
            =&\frac{1}{4\beta^{2}\sqrt{(\beta-1)^{2}+8}}\left(\beta-(2(\beta-1)^{2}+16+2\sqrt{(\beta-1)^{2}+8})\right)\\
            =&\frac{1}{4\beta^{2}\sqrt{(\beta-1)^{2}+8}}\left(-2\beta^{2}+3\beta-2-16-2\sqrt{(\beta-1)^{2}+8}\right)\\
            <&\frac{1}{4\beta^{2}\sqrt{(\beta-1)^{2}+8}}\left(-2\beta^{2}+3\beta-18-2(\beta-1)\right)\\
            =&\frac{1}{4\beta^{2}\sqrt{(\beta-1)^{2}+8}}\left(-2\beta^{2}+\beta-16\right)\\
            =&\frac{1}{4\beta^{2}\sqrt{(\beta-1)^{2}+8}}\left(-\beta(2\beta-1)-16\right)<0
        \end{split}
    \end{equation*}

\medskip

\noindent{\bf Proof of Proposition~\ref{prop:qandlambda}: }
We first show that $\epsilon_{13}^{*}$ decreases in $q$. Note that the left hand side of Equation~\eqref{eq:onecandidateepsilon}, denoted as $\mathscr{G}_{13}$, decreases in both $q$ and $\epsilon_{13}^{*}$ because $G$ decreases in both $q$ and $\epsilon_{13}^{*}$. Thus,
\begin{equation*}
    \begin{split}
        \frac{\partial \mathscr{G}_{13}}{\partial \epsilon_{13}^{*}}\frac{\partial\epsilon_{13}^{*}}{\partial q}+\frac{\partial \mathscr{G}_{13}}{\partial q}&=0\\
        \frac{\partial\epsilon_{13}^{*}}{\partial q}&=-\frac{\partial \mathscr{G}_{13}}{\partial q}\left(\frac{\partial \mathscr{G}_{13}}{\partial \epsilon_{13}^{*}}\right)^{-1}<0.
    \end{split}
\end{equation*}
Next, we show that $\epsilon_{13}^{*}$ increases in $\lambda$. To see that, first note that we must have
\begin{equation*}
    \begin{split}
        t(1-2\frac{3}{4})-\epsilon_{13}^{*}-q&<0\\
        \epsilon_{13}^{*}&>-\frac{t}{2}-q.
    \end{split}
\end{equation*}
Otherwise
\begin{equation*}
    \begin{split}
         \int_{0}^{\frac{1}{2}}G\left((t(1-2x_{n})-\epsilon_{13}^{*}-q)\frac{\sqrt{\sigma^{2}_{Q}+\beta\sigma_{s}^{2}}}{\sqrt{2}\sigma_{Q}^{2}}\right)\,dx_{n}&>\frac{1}{2}\\
         \int^{1}_{\frac{1}{2}}G\left((t(1-2x_{n})-\epsilon_{13}^{*}-q)\left(\frac{\sigma_{Q}^{4}}{\sigma_{Q}^{2}+\lambda\beta\sigma_{s}^{2}}+\frac{\sigma_{Q}^{4}}{\sigma_{Q}^{2}+\beta\sigma_{s}^{2}}\right)^{-\frac{1}{2}}\right)\,dx_{n}&\geq\frac{1}{2}
    \end{split}
\end{equation*}
which violates Equation~\eqref{eq:onecandidateepsilon}. Next we prove $\frac{\partial \mathscr{G}_{13}}{\partial \lambda}<0$. First, if $t(1-2x_{n})-\epsilon_{13}^{*}-q\leq 0$ for all $x_{n}\in[\frac{1}{2},1]$, it is obviously true as $\left(\frac{\sigma_{Q}^{4}}{\sigma_{Q}^{2}+\lambda\beta\sigma_{s}^{2}}+\frac{\sigma_{Q}^{4}}{\sigma_{Q}^{2}+\beta\sigma_{s}^{2}}\right)^{-\frac{1}{2}}$ increases in $\lambda$. Next, if $t(1-2x_{n})-\epsilon_{13}^{*}-q> 0$ for some $x_{n}\in[\frac{1}{2},1]$, there must exist some $\bar{x}\in[\frac{1}{2},\frac{3}{4}]$ such that $t(1-2x_{n})-\epsilon_{13}^{*}-q> 0$ if and only if $x_{n}<\bar{x}$, then we have
\begin{equation*}
    \begin{split}
        &\int^{1}_{\frac{1}{2}}G\left((t(1-2x_{n})-\epsilon_{13}^{*}-q)\left(\frac{\sigma_{Q}^{4}}{\sigma_{Q}^{2}+\lambda\beta\sigma_{s}^{2}}+\frac{\sigma_{Q}^{4}}{\sigma_{Q}^{2}+\beta\sigma_{s}^{2}}\right)^{-\frac{1}{2}}\right)\,dx_{n}\\
        =&\int^{\frac{1}{2}+2(\bar{x}-\frac{1}{2})}_{\frac{1}{2}}G\left((t(1-2x_{n})-\epsilon_{13}^{*}-q)\left(\frac{\sigma_{Q}^{4}}{\sigma_{Q}^{2}+\lambda\beta\sigma_{s}^{2}}+\frac{\sigma_{Q}^{4}}{\sigma_{Q}^{2}+\beta\sigma_{s}^{2}}\right)^{-\frac{1}{2}}\right)\,dx_{n}\\
        &+\int_{\frac{1}{2}+2(\bar{x}-\frac{1}{2})}^{1}G\left((t(1-2x_{n})-\epsilon_{13}^{*}-q)\left(\frac{\sigma_{Q}^{4}}{\sigma_{Q}^{2}+\lambda\beta\sigma_{s}^{2}}+\frac{\sigma_{Q}^{4}}{\sigma_{Q}^{2}+\beta\sigma_{s}^{2}}\right)^{-\frac{1}{2}}\right)\,dx_{n}\\
        =&\frac{1}{2}+\int_{\frac{1}{2}+2(\bar{x}-\frac{1}{2})}^{1}G\left((t(1-2x_{n})-\epsilon_{13}^{*}-q)\left(\frac{\sigma_{Q}^{4}}{\sigma_{Q}^{2}+\lambda\beta\sigma_{s}^{2}}+\frac{\sigma_{Q}^{4}}{\sigma_{Q}^{2}+\beta\sigma_{s}^{2}}\right)^{-\frac{1}{2}}\right)\,dx_{n}
    \end{split}
\end{equation*}
which is increasing in $\lambda$ as $(t(1-2x_{n})-\epsilon_{13}^{*}-q<0$ for all $x_{n}\geq\frac{1}{2}+2(\bar{x}-\frac{1}{2})$. Thus, $\frac{\partial \mathscr{G}_{13}}{\partial \lambda}<0$, and
\begin{equation*}
    \begin{split}
        \frac{\partial \mathscr{G}_{13}}{\partial \epsilon_{13}^{*}}\frac{\partial\epsilon_{13}^{*}}{\partial \lambda}+\frac{\partial \mathscr{G}_{13}}{\partial \lambda}&=0\\
        \frac{\partial\epsilon_{13}^{*}}{\partial \lambda}&=-\frac{\partial \mathscr{G}_{13}}{\partial \lambda}\left(\frac{\partial \mathscr{G}_{13}}{\partial \epsilon_{13}^{*}}\right)^{-1}<0.
    \end{split}
\end{equation*}
The results regarding $\epsilon_{123}^{*}$ follow similar arguments.

Now for the overall effect, as shown above, when $q$ or $\lambda$ increase, both $\epsilon_{13}^{*}$ and $\epsilon_{123}^{*}$ decreases. Thus $\frac{1}{2}G\left(\frac{\epsilon_{123}^{*}}{\sigma_{\epsilon}}\right)$ and $G\left(\frac{\epsilon_{13}^{*}}{\sigma_{\epsilon}}\right)$ decreases, and the number of challengers decrease in $q$ and $\lambda$. When Equation~\eqref{eq:propmorechallengers} does not hold, as shown in Proposition~\ref{prop:morechallengers}, a decrease in the number of challengers necessarily imply that incumbent is re-elected with a higher probability. Thus, combined with the fact that $\frac{1}{2}G\left(\frac{\epsilon_{123}^{*}}{\sigma_{\epsilon}}\right)$ and $G\left(\frac{\epsilon_{13}^{*}}{\sigma_{\epsilon}}\right)$ decreases in $q$ and $\lambda$, the probability of the incumbent being re-elected also decreases in $q$ and $\lambda$.

To show that the overall effect is non-monotonic when Equation~\eqref{eq:propmorechallengers} holds, consider the example where $C=\frac{1}{2}G\left(\frac{\epsilon_{123}^{*}}{\sigma_{\epsilon}}\right)<G\left(\frac{\epsilon_{13}^{*}}{\sigma_{\epsilon}}\right)$. In this scenario, two challengers enter the race. Now consider a small increase in $q$ or $\lambda$, it implies that $G\left(\frac{\epsilon_{13}^{*}}{\sigma_{\epsilon}}\right)>C>\frac{1}{2}G\left(\frac{\epsilon_{123}^{*}}{\sigma_{\epsilon}}\right)$, and now only one challenger enters the race. By Proposition~\ref{prop:morechallengers}, the incumbent is re-elected with a lower probability when the change in $q$ and $\lambda$ is small enough.


\medskip 

\noindent {\bf Proof of Corollary \ref{coro:morechallengers}:}
We divide the proof into two parts. First, we prove that $\frac{\partial \epsilon_{123}^{*}}{\partial \sigma_{s}^{2}}\Bigl\vert_{\beta\sigma_{s}^{2}})$ is positive if inequality~\eqref{eq:propmorechallengers} holds. It implies that keeping $\beta\sigma_{s}^{2}$ fixed, a marginal drop in $\sigma_{s}^{2}$ decreases $\epsilon_{123}^{*}$, i.e., increases incumbency advantage. Second, we prove that fixing $\beta\sigma_{s}^{2}$, if $\frac{\partial \epsilon_{123}^{*}}{\partial \sigma_{s}^{2}}\Bigl\vert_{\beta\sigma_{s}^{2}}<0$ for some $\sigma_{s}^{2}$, we also have $\frac{\partial \epsilon_{123}^{*}}{\partial \sigma_{s}^{2}}\Bigl\vert_{\beta\sigma_{s}^{2}}<0$ for some $\tilde{\sigma}_{s}^{2}<\sigma_{s}^{2}$, which proves the corollary.

Denote that the left hand side of equation~\eqref{eq:twocandidateepsilon} as $\mathscr{G}_{123}$. Note that as $\frac{\partial \epsilon_{123}^{*}}{\partial \sigma_{s}^{2}}\Bigl\vert_{\beta\sigma_{s}^{2}}=-\left(\frac{\partial \mathscr{G}_{123}}{\partial \epsilon_{123}^{*}}\right)^{-1}\left(\frac{\partial \mathscr{G}_{123}}{\partial \sigma_{s}^{2}}\Bigl\vert_{\beta\sigma_{s}^{2}}\right)$ and $\left(\frac{\partial \mathscr{G}_{123}}{\partial \epsilon_{123}^{*}}\right)^{-1}<0$, we prove that the left hand side of equation~\eqref{eq:twocandidateepsilon}, denoted as $\mathscr{G}_{123}$, increases in $\sigma_{s}^{2}$ when $\beta\sigma_{s}^{2}$ is fixed. 
First, when $\beta\sigma_{s}^{2}$ is fixed, the second item of $\mathscr{G}_{123}$ is unchanged. Second, as equation~\eqref{eq:twocandidateepsilon} implies that $t(1-\frac{1}{4})-\epsilon_{123}^{*}-q>0$, fixing $\beta\sigma_{s}^{2}$, $\mathscr{G}_{123}$ increases in $\sigma_{s}^{2}$ if and only if
\begin{equation*}
    \begin{split}
        \frac{\partial}{\partial\sigma_{s}^{2}}\left(\frac{\sigma_{Q}^{4}}{\sigma_{Q}^{2}+\beta\sigma_{s}^{2}}+\frac{\frac{\sigma_{Q}^{2}+\beta\sigma_{s}^{2}}{\beta^{2}\sigma_{s}^{4}}}{(\frac{1}{\sigma_{Q}^{2}}+\frac{1}{\beta\sigma_{s}^{2}}+\frac{1}{\sigma_{s}^{2}})^{2}}+\frac{\frac{\sigma_{Q}^{2}+\sigma_{s}^{2}}{\sigma_{s}^{4}}}{(\frac{1}{\sigma_{Q}^{2}}+\frac{1}{\beta\sigma_{s}^{2}}+\frac{1}{\sigma_{s}^{2}})^{2}}\right)\Bigl\vert_{\beta\sigma_{s}^{2}}<0.
    \end{split}
\end{equation*}
With simple algebra,
\begin{equation*}
    \begin{split}
        &\frac{\partial}{\partial\sigma_{s}^{2}}\left(\frac{\sigma_{Q}^{4}}{\sigma_{Q}^{2}+\beta\sigma_{s}^{2}}+\frac{\frac{\sigma_{Q}^{2}+\beta\sigma_{s}^{2}}{\beta^{2}\sigma_{s}^{4}}}{(\frac{1}{\sigma_{Q}^{2}}+\frac{1}{\beta\sigma_{s}^{2}}+\frac{1}{\sigma_{s}^{2}})^{2}}+\frac{\frac{\sigma_{Q}^{2}+\sigma_{s}^{2}}{\sigma_{s}^{4}}}{(\frac{1}{\sigma_{Q}^{2}}+\frac{1}{\beta\sigma_{s}^{2}}+\frac{1}{\sigma_{s}^{2}})^{2}}\right)\Bigl\vert_{\beta\sigma_{s}^{2}}\\
        =&2\frac{\frac{\sigma_{Q}^{2}+\beta\sigma_{s}^{2}}{\beta^{2}\sigma_{s}^{4}}}{(\frac{1}{\sigma_{Q}^{2}}+\frac{1}{\beta\sigma_{s}^{2}}+\frac{1}{\sigma_{s}^{2}})^{3}}(\frac{1}{\sigma^{4}_{s}})+\frac{-\frac{2\sigma_{Q}^{2}}{\sigma_{s}^{6}}-\frac{1}{\sigma_{s}^{4}}}{(\frac{1}{\sigma_{Q}^{2}}+\frac{1}{\beta\sigma_{s}^{2}}+\frac{1}{\sigma_{s}^{2}})^{2}}+2\frac{\frac{\sigma_{Q}^{2}+\sigma_{s}^{2}}{\sigma_{s}^{4}}}{(\frac{1}{\sigma_{Q}^{2}}+\frac{1}{\beta\sigma_{s}^{2}}+\frac{1}{\sigma_{s}^{2}})^{3}}(\frac{1}{\sigma^{4}_{s}})
    \end{split}
\end{equation*}
which is smaller than $0$ if and only if
\begin{equation}
    \begin{split}
        2\frac{\sigma_{Q}^{2}+\beta\sigma_{s}^{2}}{\beta^{2}\sigma_{s}^{4}}+2\frac{\sigma_{Q}^{2}+\sigma_{s}^{2}}{\sigma_{s}^{4}}&<(\frac{2\sigma_{Q}^{2}}{\sigma_{s}^{2}}+1)(\frac{1}{\sigma_{Q}^{2}}+\frac{1}{\beta\sigma_{s}^{2}}+\frac{1}{\sigma_{s}^{2}})\\
        2(\sigma_{Q}^{2}+\beta\sigma_{s}^{2})+2\beta^{2}(\sigma_{Q}^{2}+\sigma_{s}^{2})&<(2\beta^{2}\sigma_{Q}^{2}\sigma_{s}^{2}+\beta^{2}\sigma_{s}^{4})(\frac{1}{\sigma_{Q}^{2}}+\frac{1}{\beta\sigma_{s}^{2}}+\frac{1}{\sigma_{s}^{2}})\\
        2(\sigma_{Q}^{2}+\beta\sigma_{s}^{2})+2\beta^{2}(\sigma_{Q}^{2}+\sigma_{s}^{2})&<2\beta^{2}\sigma_{s}^{2}+2\beta\sigma_{Q}^{2}+2\beta^{2}\sigma_{Q}^{2}+\beta^{2}\frac{\sigma_{s}^{4}}{\sigma_{Q}^{2}}+\beta\sigma_{s}^{2}+\beta^{2}\sigma_{s}^{2}\\
        2+2\beta^{2}(2\beta+2\beta^{2})\frac{\sigma_{s}^{2}}{\sigma_{Q}^{2}}&<\beta^{2}\left(\frac{\sigma_{s}^{2}}{\sigma_{Q}^{2}}\right)^{2}+(3\beta^{2}+\beta)\frac{\sigma_{s}^{2}}{\sigma_{Q}^{2}}+2\beta+2\beta^{2}\\
        \beta^{2}\left(\frac{\sigma_{s}^{2}}{\sigma_{Q}^{2}}\right)^{2}+\beta(\beta-1)\frac{\sigma_{s}^{2}}{\sigma_{Q}^{2}}&>2(1-\beta)
    \end{split}
    \label{eq:morethan2challengers}
\end{equation}
which holds if inequality~\eqref{eq:propmorechallengers} holds because $2(1-\beta)<2$. We now prove that fixing $\beta\sigma_{s}^{2}$, if $\frac{\partial \epsilon_{123}^{*}}{\partial \sigma_{s}^{2}}\Bigl\vert_{\beta\sigma_{s}^{2}}<0$ for some $\sigma_{s}^{2}$, we also have $\frac{\partial \epsilon_{123}^{*}}{\partial \sigma_{s}^{2}}\Bigl\vert_{\beta\sigma_{s}^{2}}<0$ for some $\tilde{\sigma}_{s}^{2}<\sigma_{s}^{2}$. As $\beta\sigma_{s}^{2}$ is fixed, a smaller $\sigma_{s}^{2}$ implies a bigger $\beta$. First note that the inequality~\eqref{eq:morethan2challengers} holds when $\beta\geq 1$. Next, when $\beta<1$, inequality~\eqref{eq:morethan2challengers} can be rewritten as:
\begin{equation*}
    \begin{split}
        \beta^{2}\left(\frac{\sigma_{s}^{2}}{\sigma_{Q}^{2}}\right)^{2}+\beta(\beta-1)\frac{\sigma_{s}^{2}}{\sigma_{Q}^{2}}&>2(1-\beta)\\
        \left(\frac{\beta\sigma_{s}^{2}}{\sigma_{Q}^{2}}+\frac{\beta-1}{2}\right)^{2}&>2(1-\beta)\\
        \frac{\beta\sigma_{s}^{2}}{\sigma_{Q}^{2}}&>\sqrt{2(1-\beta)}-\frac{\beta-1}{2}
    \end{split}
\end{equation*}
where the right-hand side is increasing in $\beta$. The result thus follows.

}

\subsection*{APPENDIX B: {\small Micro-foundation for Challengers and Information Provision}}
In this section, we show a simple model to illustrate a channel that induces more information when more challengers enter the race.

Suppose there are $N$ candidates in the primary, and each decides how many signals to send to the voters. Assume for simplicity that they get utility $1$ if they win the primary, and that their winning probability increases in the number of signals they send compared to their competitors. This assumption is motivated by the fact that more signals make a candidate more visible and thus win with a higher probability. More specifically, denoting the number of signals sent by candidate $i$ with $q_{i}$, we assume that the winning probability of a candidate follows the Tullock contest function \citep{buchanan1980toward}, which is equals to:
\begin{equation*}
    \frac{q_{i}^{r}}{\sum_{j=1}^{N}q_{j}^{r}}
\end{equation*}
where $r\leq 1$. We also assume that the cost of the number of signals follows a quadratic form $\frac{A}{2}q_{i}^{2}$ for some constant $A$. The Tullock contest function ensures that the winning probability is increasing and concave in $q_{i}$, such that the best response is characterized by the first order condition.

In the following, we derive the equilibrium number of signals, denoted as $q^{*}$. In a symmetric equilibrium, given all candidates other than $i$ send $q^{*}$ signals, the best response of candidate $i$ equals to $q^{*}$ and follows the following first order condition:
\begin{equation*}
    \begin{split}
        \frac{\left(\sum_{j\neq i}(q^{*}){r}\right)r(q^{*})^{r-1}}{\left(\sum_{j=1}^{N}(q^{*})^{r}\right)^{2}}&=Aq^{*}\\
        \frac{r(N-1)(q^{*})^{2r-1}}{N^{2}(q^{*})^{2r}}&=Aq^{*}\\
        (q^{*})^{2}&=\frac{r(N-1)}{AN}
    \end{split}
\end{equation*}
which clearly shows that $q^{*}$ increases in $N$. The intuition of this result is reminiscent to the literature of advertising. In particular, as there are more candidates, the aggregate signals in the whole market increases, and thus it is more difficult for a particular candidate to stand out, and in equilibrium each candidate invests more on information provision.

}

\end{document}